\documentclass[11pt,letter]{article}
\usepackage[dvips]{epsfig} 
\usepackage{graphicx,color}
\usepackage{hyperref,setspace,amssymb,comment}
\newcommand{\email}[1]{%
  \href{mailto:#1}{\footnotesize\nolinkurl{#1}}}

\newcommand{\farcs}{\mbox{\ensuremath{.\!\!^{\prime\prime}}}}%  % fractional arcsecond symbol: 0.''0
\newcommand{\arcsec}{\mbox{\ensuremath{^{\prime\prime}}}}

\def\kms  {km~s$^{-1}$}

\def\uasy {$\mu$as~y$^{-1}$}

\def\uJy  {$\mu$Jy}

\def\Vlsr {\ifmmode {V_{\rm LSR}}\else {$V_{\rm LSR}$}\fi}
\def\Msun {\ifmmode {M_\odot}\else {$M_\odot$}\fi}
\def\Ro   {\ifmmode {R_0}\else {$R_0$}\fi}
\def\Ho   {\ifmmode {H_0}\else {$H_0$}\fi}
\def\To   {\ifmmode {\Theta_0}\else {$\Theta_0$}\fi}
\def\Vsun {\ifmmode {V_\odot}\else {$V_\odot$}\fi}

\newcommand{\msun}{\mbox{$M_{\odot}$}~}
\def\sgra{Sgr~A$^*$}
\def\mdot{\dot{\rm M}}

\begin{document}

\begin{center}
\includegraphics[width=\textwidth]{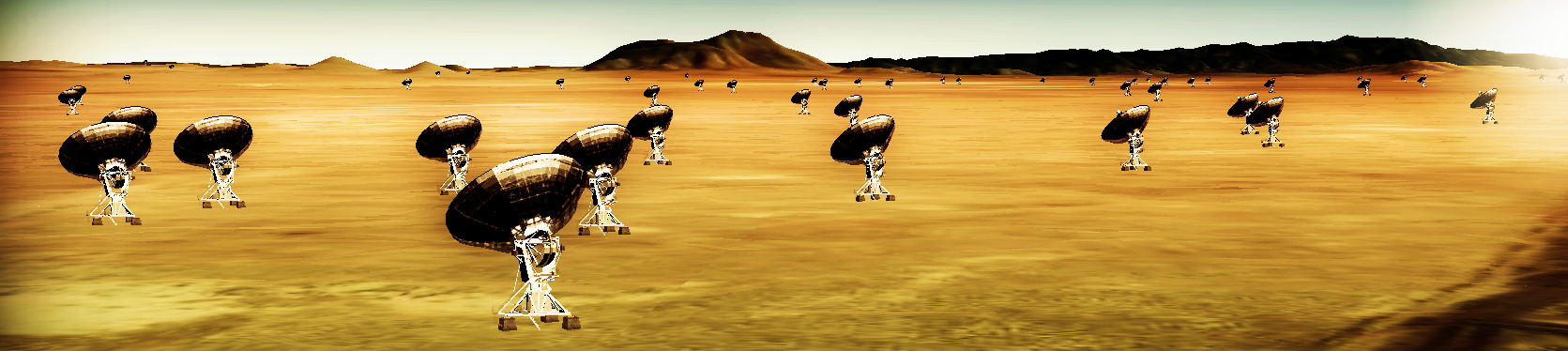}
\end{center}

\begin{center}

{\bf \Large Next Generation Very Large Array Memo No. 7}

\vspace{0.1in}

{\bf \Large Science Working Group 2}

\vspace{0.1in}

{\bf \Large ``Galaxy Ecosystems'' : The Matter Cycle in and Around Galaxies}

\end{center}

\hrule

\vspace{0.7cm}

\noindent
Adam K. Leroy,$^1$ Eric Murphy,$^2$ Lee Armus,$^2$ Crystal Brogan,$^3$ Jennifer Donovan Meyer,$^3$ Aaron Evans,$^{3,4}$ Todd Hunter$^3$, Kelsey Johnson,$^4$ Jin Koda$^5$, David S. Meier,$^6$ Karl Menten,$^7$ Elizabeth Mills,$^8$ Emmanuel Momjian,$^{8}$ Juergen Ott,$^8$ Frazer Owen,$^8$ Mark Reid,$^9$ Erik Rosolowsky$^{10}$, Eva Schinnerer$^{11}$, Nicholas Scoville,$^{12}$Kristine Spekkens,$^{13}$ Liese van Zee,$^{14}$ Tony Wong$^{15}$\\

\begin{center}
{\bf \large Abstract}\\
\end{center}

This white paper discusses how a ``next-generation" Very Large Array (ngVLA) operating in the frequency range 1--116~GHz could be a groundbreaking tool to study the detailed astrophysics of the ``matter cycle'' in the Milky Way and other galaxies. If optimized for high brightness sensitivity, the ngVLA would bring detailed microwave spectroscopy and modeling of the full radio spectral energy distribution into regular use as survey tools at resolution $0.1$--$1\arcsec$. This wavelength range includes powerful diagnostics of density, excitation, and chemistry in the cold ISM, as well as multiple tracers of the rate of recent star formation, the magnetic field, shocks, and properties of the ionized ISM. We highlight design considerations that would make this facility revolutionary in this area, the foremost of which is a large amount of collecting area on $\sim$km-length baselines. We also emphasize the strong case for harnessing the large proposed collecting area of the ngVLA for very long baseline applications as part of the core design. This would allow measurements of the three dimensional space motions of galaxies to beyond the Local Group and mapping of the Milky Way out to the far side of the disk. Finally, we discuss the gains from the proposed combination of very high resolution and sensitivity to thermal emission, which include observing the feeding of black holes and resolving forming protoclusters.

\vspace{2mm}

\begin{spacing}{1}
\footnotesize
\noindent $^1$The Ohio State University, 140 W 18th St, Columbus, OH 43210, USA; \email{leroy.42@osu.edu}\\
\noindent $^2$Infrared Processing and Analysis Center, California Institute of Technology, MC 220-6, Pasadena, CA 91125, USA; \email{emurphy@ipac.caltech.edu}\\
\noindent $^3$National Radio Astronomy Observatory, 520 Edgemont Rd, Charlottesville, VA 22903, USA\\
\noindent $^4$Department of Astronomy, University of Virginia, Charlottesville, VA 22904, USA\\
\noindent $^5$Department of Physics and Astronomy, Stony Brook University, Stony Brook, NY 11794-3800, USA\\
\noindent $^6$ New Mexico Institute of Mining and Technology, 801 Leroy Place, Socorro, NM 87801, USA\\
\noindent $^7$ Max-Planck-Institut f\"ur Radioastronomie, Auf dem H\"ugel 69, D-53121 Bonn, Germany\\
\noindent $^8$National Radio Astronomy Observatory, P.O. Box O, 1003 Lopezville Road, Socorro, NM 87801, USA\\
\noindent $^9$Harvard-Smithsonian Center for Astrophysics, 60 Garden Street, Cambridge, MA 02138, USA\\
\noindent $^{10}$Department of Physics, University of Alberta, Edmonton, AB, Canada\\
\noindent $^{11}$Max Planck Institut f\"{u}r Astronomie, K\"{o}nigstuhl 17, Heidelberg D-69117, Germany\\
\noindent $^{12}$California Institute of Technology, MC 249-17, 1200 East California Boulevard, Pasadena, CA 91125, USA\\
\noindent $^{13}$Department of Physics, Royal Military College of Canada, PO Box 17000, Station Forces, Kingston, Ontario, Canada, K7K 7B4\\
\noindent $^{14}$Department of Astronomy, Indiana University, 727 E 3rd St, Bloomington, IN 47405, USA\\
\noindent $^{15}$Department of Astronomy, University of Illinois, Urbana, IL 61801, USA\\
\end{spacing}

\newpage

\tableofcontents

\newpage

\begin{comment}
\begin{picture}(30,100)
\vspace{-1cm} 
\put(23,145){  {\bf \large \kern1.9cmNext Generation VLA Science White Paper }}
\put(20,125){ \small \kern1.98cm Working Group 2: ``Galaxy Ecosystems'' and the Matter Cycle in Galaxies}

\end{picture}
\vspace{-3.8cm}
\hrule width 14.3cm
\vspace{0.2cm}

\vspace{1cm}
\begin{figure}
\epsfig{file=NRAO_logo.eps,height=1.5cm} 
\end{figure}

\vspace{-0.6cm}

\begin{flushright}
\today
\end{flushright}

\vspace{1cm}

\begin{abstract}
\end{abstract}
\end{comment}

\section{Overview}

Radio telescopes are essential to our understanding of galaxies as ``ecosytems,'' meaning the complex interplay of neutral and ionized gas, stars, black holes, dark matter, and magnetic fields in and around galaxies. They represent a main tool to study the baryon cycle by providing access to almost all phases of gas in galaxies: cool and cold gas (via emission and absorption lines), ionized gas (via free-free continuum and recombination lines), cosmic rays and hot gas (via synchrotron emission and the Sunyaev-Zeldovich effect). These telescopes make it possible to observe these phases free from the dust extinction which plagues shorter wavelengths. Modern radio interferometers allow one to measure a wide range of angular scales and to observe with exquisite frequency resolution over a wide bandwidth. This makes them excellent tools to study the motions of gas from the immediate vicinity of black holes and forming stars out to the dark-matter dominated halos of galaxies. The ability of radio interferometry to work across very long baselines also gives radio telescopes high resolving power, allowing one to directly observe the motions of local galaxies, forming stars, and material orbiting supermassive black holes. Over the previous decades, the Very Large Array (VLA) and the Very Long Baseline Array (VLBA) have made fundamental contributions to each of these areas via observations at $\sim$cm wavelengths.  Additionally, arrays like the Combined Array for mm-Wave Astronomy (CARMA) and its predecessors,  the Plateau de Bure Interferometer (PdBI), and the Nobeyama Millimeter Array (NMA) have done so at $\sim$ few-mm wavelengths. The Atacama Large Millimeter/submillimeter Array (ALMA) is poised to make great strides in sun-mm astrophysics in the coming decades.

A ``next-generation Very Large Array'' (ngVLA) focused on the rich frequency range $\nu \sim 1$--$116$~GHz has been proposed as a next major direction for the U.S. astronomical community (see the associated overview white paper). In this document, we describe how, if properly designed, such a facility could allow major breakthroughs in the area of ``galaxy ecosystems.'' Through a rich set of spectral and continuum diagnostics, this part of the spectrum offers access to the physics of almost all phases of interstellar matter cycling. However, beyond a few bright lines and pioneering detailed studies of individual bright sources, the full power of cm- and mm-wave observations remains largely untapped. The fundamental limitation has so far been the surface brightness sensitivity of previous- and current-generation cm- and mm-wave arrays. The emission from tracers that can provide accurate measurements of the gas and plasma properties in galaxies has simply been too faint to reliably use as a general purpose tool. 

\begin{figure}[t!]
\centering
\includegraphics[width=\textwidth]{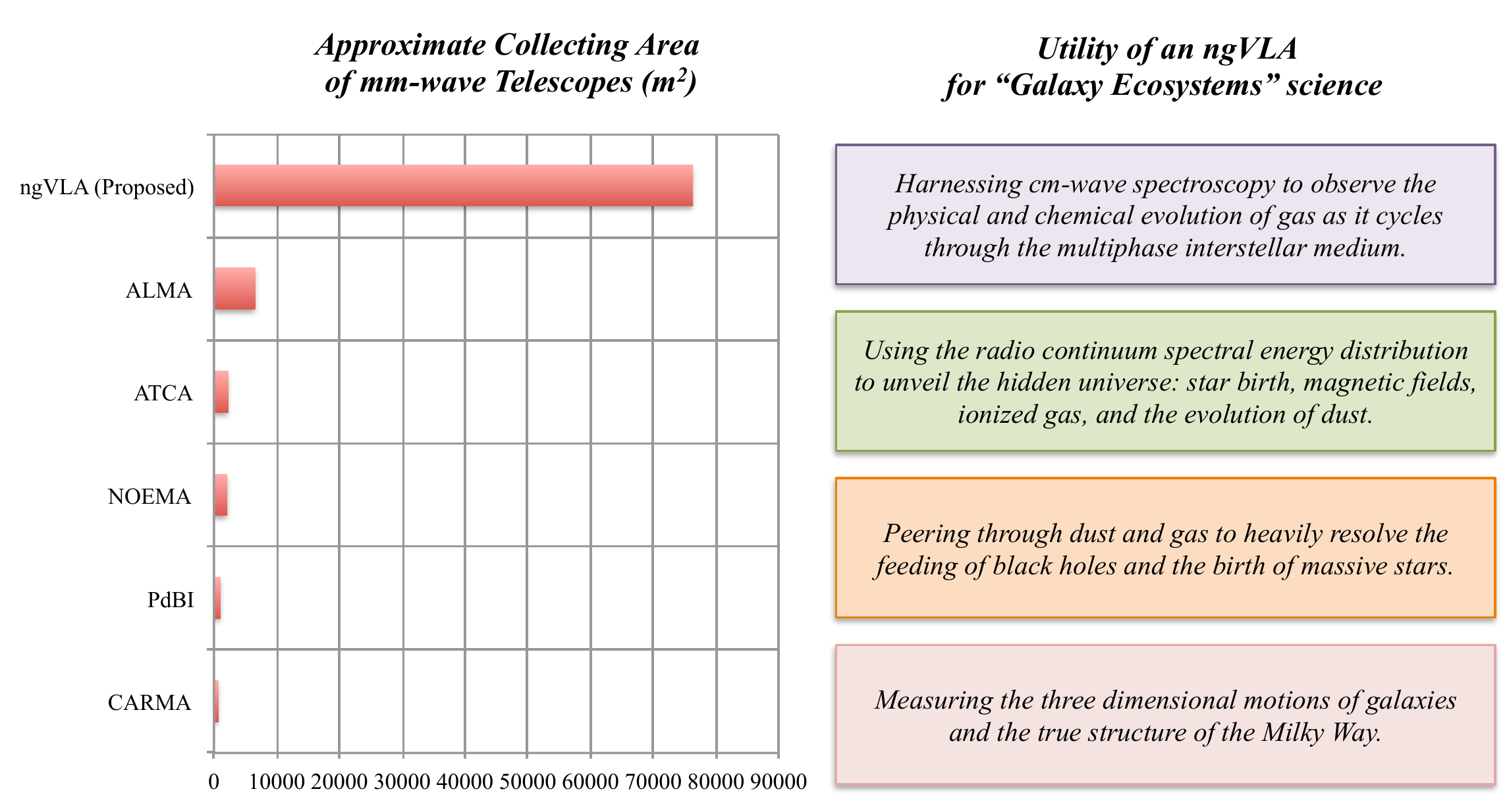}
\caption{\small {\bf ({\em Left}):} Visual comparison of proposed collecting area for the ngVLA and values for current-generation mm-wave interferometers. The proposed ngVLA represents an order of magnitude improvement over current-generation facilities operating in the diagnostic-rich $\lambda \sim 3$--$10$~mm range, ensuring the revolutionary nature of such a facility. ({\bf {\em Right}:}) Major themes for advancement in the area of ``Galaxy Ecosystems'' that would be enabled by such a facility.}
\end{figure}

The ngVLA under discussion could change this situation, bringing a large suite of new diagnostics into general use. From the perspective of galaxy ecosystems, the major advance with the ngVLA would be the enormous gain in sensitivity at many spatial scales due to an unprecedented collecting area operating in the $\nu \sim 1$--$116$~GHz range. This frequency range includes irreplaceable low-excitation molecular transitions and continuum emission from all major long-wavelength emission processes. These spectral diagnostics are the key to understanding the chemistry and physics of the interstellar material in and around galaxies. The continuum diagnostics offer the chance to study the ionized gas, dust, cosmic rays, and magnetic fields in galaxies (including our own) in ways not possible at other wavelengths.

These tools have been recognized for decades and explored in narrow contexts by pushing the capabilities of previous-generation facilities. However, the limited sensitivity of mm- and cm-wave facilities has so far prevented them from becoming general purpose tools. The linear scaling of a telescope's sensitivity with collecting area (compared to the weaker $\sigma \propto \Delta t^{-0.5} \Delta \nu^{-0.5}$ dependence on integration time or bandwidth) means that the key to bringing these diagnostics into the mainstream must be a large area interferometer. Even for applications focused on resolution, sensitivity --- driven by collecting area --- is often the limiting factor thanks to the non-linear interplay of resolution and surface brightness sensitivity for an interferometer. In this sense, imagining the ngVLA as a large collecting area mm- and cm-wave facility makes it a natural complement and a natural next step to other recent long-wavelength facilities. The collecting area under discussion exceeds that of the Atacama Large Millimeter/submillimeter Array (ALMA) and the currently-under-construction Northern Extended Millimeter Array (NOEMA) by an order of magnitude. This simple metric (collecting area) and the irreplaceable diagnostic capabilities available only in the $\lambda \sim 3$--$10$~mm window mean that the ngVLA has the potential to have a large impact even when constructed a decade or more from now. For an order of magnitude difference in collecting area, previous facilities would need to integrate for $\sim 100$ years to match the output of a single year of an ngVLA at matched frequency and bandwidth. This basic consideration illustrates that this kind of facility is a logical ``next generation'' goal for the U.S. community with the potential to be groundbreaking when constructed even given the fast-moving pace of modern astronomy.

As with ALMA or the upgraded Karl G. Janksy VLA, the ngVLA would be a stunningly flexible facility. Its practical application to the area of ``galaxy ecosystems'' would be to carry out an enormous breadth of principle investigator (PI) driven science, executing programs that amount to ``key projects'' with current facilities in only a few hours and using larger allocations to run experiments that are currently impossible. The goal of this white paper is to illustrate the breadth of science that makes a facility like this in this frequency range so exciting. Doing so, we hope to capture the imaginations of, and start a conversation among, interested parties that will shape the goals and design of such a telescope in the coming years. 

We also aim to highlight aspects of telescope design that will be important to optimize an ngVLA as a facility to study the ``baryon cycle'' or ``galaxy ecoystems.'' For two of the four main areas in this white paper, one item stands out as absolutely crucial: {\em the ngVLA must dramatically exceed the effective collecting area of ALMA on the $\sim$ km-length baselines relevant to spectroscopic imaging of the cold ISM and studies of faint continuum emission from normal galaxies.} If this goal is met, the ngVLA has the prospect to open up a huge range of spectroscopic and continuum diagnostics for use in surveys of the Milky Way and other galaxies. The faintness of these diagnostics mean that their true potential is likely to be unlocked only with an observatory with significantly enhanced capabilities compared to ALMA. The ngVLA is a natural way to envision taking this next step. However, note that this goal is not trivial: the notional ngVLA as presented in the introduction does not yet meet this requirement, instead placing most of the impressive proposed collecting area at longer baselines. As the ngVLA concept is discussed in the coming years, the prospective gain from various design considerations will need to be carefully weighed. Doing so, ALMA's performance in its ``Bands 1, 2, and 3'' --- along with likely upgrades --- must represent the logical point of comparison.

A number of additional design considerations also emerge from considering potential gains in the area of galaxy ecosystems.

\begin{itemize}
\item A {\bf wide instantaneous bandwidth} and the ability to {\bf heavily multiplex spectral line and continuum observations}. A wide bandwidth is needed for continuum sensitivity to measure the shape of the continuum spectral energy distribution, which is key to its interpretation. The diagnostic and astrochemical utility of spectral line observations is vastly improved by simultaneously observing a wide range of transitions; in fact, the comparative behavior of multiple lines and the continuum is the central measurement for this kind of science. The ability to observe a large spectral range (e.g., $75$--$116$~GHz or $20$--$50$~GHz) simultaneously with the frequency resolution to capture individual transitions will be also be a key capability. By observing many faint signals at once, one gains a multiplicative improvement in science per hour.

\item The ability to reconstruct {\bf high fidelity, full flux-recovery} maps of celestial objects is crucial and should be included in the baseline design of the telescope. ALMA achieves this through inclusion of multiple dish sizes and single-dish capabilities in its array. Other strategies are viable, but the ability to capture the full intensity distribution from the sky and to accurately reconstruct images is essential to the ``galaxy ecosystem'' science, which by its nature captures processes with a complex spatial power spectrum across a wide range of scales.

\item Coverage of the {\bf full frequency range $\nu = 1$--$116$~GHz}. Many continuum and line diagnostics lie throughout this range, which is bookended by the workhorse tracers of the molecular (CO 1--0 at $\nu = 115.3$~GHz) and atomic (the {\sc Hi} 21-cm Spin-flip transition at $\nu = 1.4$~GHz) interstellar medium. Coverage of this whole range, including the CO and {\sc Hi} lines (where collecting area allows unprecedented sensitivity), will ensure that the ngVLA has the ability to sample the entire matter cycle in galaxies and ensure a phenomenal breadth of revolutionary science.

\item {\bf Integration of the large collecting area of the ngVLA with very long baseline capabilities.} The next generation of very long baseline science --- including measurements of the motions of local galaxies, black hole masses, and the structure of the Milky Way --- depends on improving sensitivity by deploying more collecting area over baselines of many 1,000s of km. The ngVLA can advance these goals by planning to integrate with other facilities to use its large collecting area in very long baseline science as  part of the core design. A more aggressive approach that could be considered would be to place a fraction of the collecting area at very long baselines as part of the ngVLA design.

\item For survey speed reasons, {\bf smaller dish sizes} are preferred as long as they do not compromise the total collecting area. Many key applications of galaxy ecosystem science involve mapping large areas of the sky and the efficiency of this mapping is increased by a larger field of view.

\end{itemize}

\begin{figure}[t!]
\centering
\includegraphics[width=\textwidth]{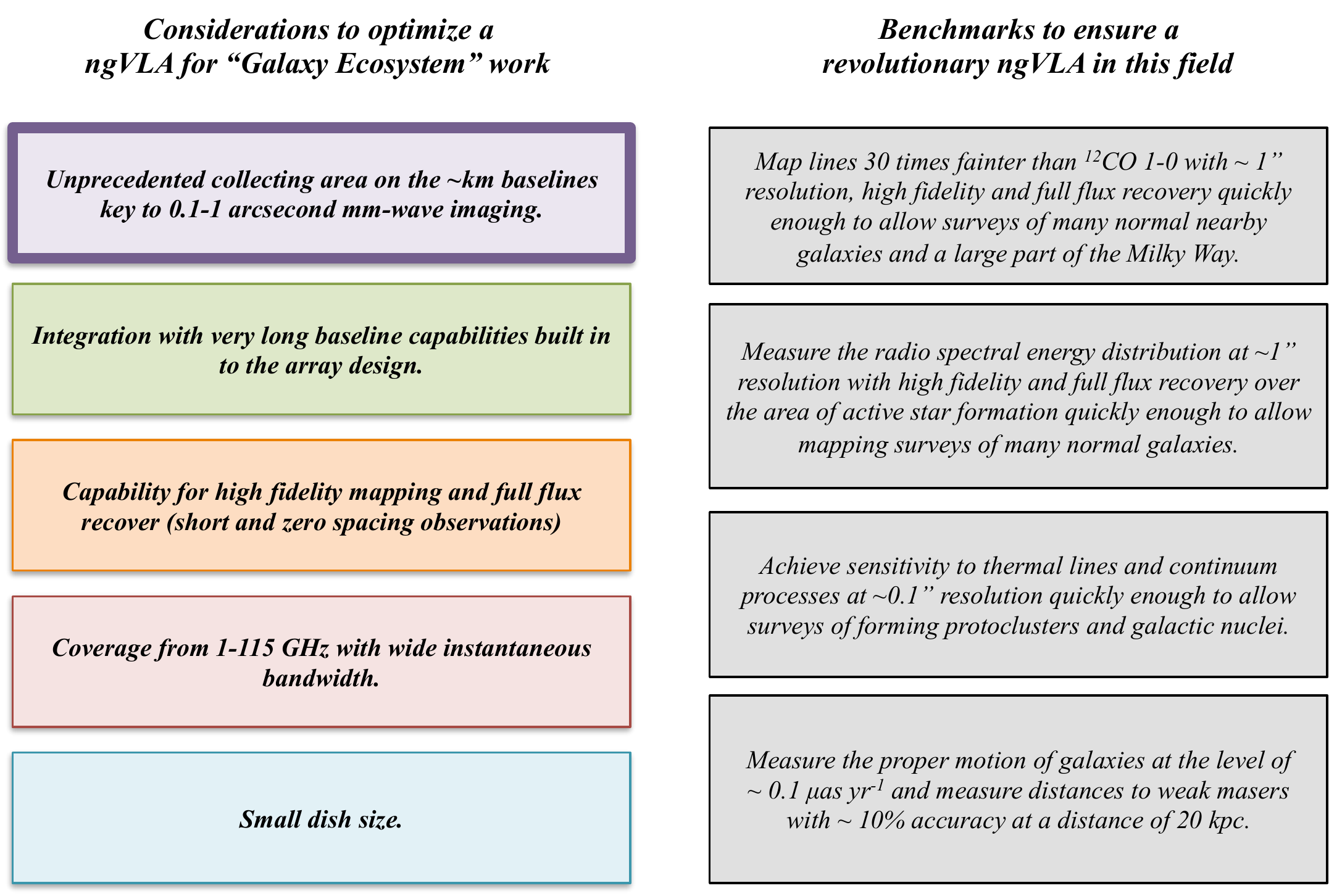}
\end{figure}

\noindent Coupling these design considerations with a large collecting area would make the ngVLA a transformational facility for studying galaxies as ecosystems. Such an instrument would open up the powerful diagnostics at cm- and mm-wavelengths, overcoming the sensitivity limitations from previous facilities. This white paper highlights the gains that we would expect in several areas related to this matter cycle in galaxies. We focus on a few themes that illustrate the new spectroscopic, resolving, and continuum capabilities that would come from marrying good resolution with high sensitivity:\\

\noindent {\bf Theme \#1:} {\em Harnessing the full potential of cm-wave spectroscopic imaging to measure the physical and chemical evolution of gas as it moves through the multiphase ISM.}\\

\noindent {\bf Theme \#2:} {\em Peering through dust and gas to resolve the feeding of supermassive black holes and the birth of high mass stars.}\\

\noindent {\bf Theme \#3:}  {\em Capturing the signatures of star formation, magnetic fields, hidden ionized gas, and dust evolution via the full radio spectral energy distribution.}\\

\noindent {\bf Theme \#4:} {\em Improving our knowledge of the Milky Way, the Local Group, supermassive black holes, and the expansion of the Universe by harnessing the collecting area of the ngVLA for very long baseline science.}

\medskip

\noindent In each section, we outline the new capabilities one would expect from the ngVLA and discuss the scientific gains from applying these capabilities to the topics of galaxy evolution since $z \sim 1$, the cycling of gas in and out of galaxies, the formation and evolution of molecular gas clouds, the birth of stars, the growth of supermassive black holes, the structure of galaxies, and feedback from stars and active galactic nuclei into the AGN.

Part of the charge to this working group was to cast design requirements in terms of a handful of ``key projects'' that could help drive a proposed telescope design. For an area as broad as ``galaxy ecosystems'' this is challenging, but we are able to highlight some concrete benchmarks that, if met, would ensure the transformational nature of the facility:

\begin{enumerate}
\item The ability to make a $1\arcsec \times 5$~km~s$^{-1}$ resolution, high fidelity, full flux recovery spectral map of a line with surface brightness $30\times$ fainter than CO (the brightness of many key diagnostic transitions) across a large nearby spiral galaxy in less than $\sim 10$--$20$~h, allowing for  spectroscopic and chemical surveys of a large set of $z = 0$ galaxies. In the same parameter space, the ability to quickly map bright thermal lines over large areas of the Milky Way or another galaxy.

\item The ability to measure the radio spectral energy distribution with high fidelity and full flux recovery at $1\arcsec$ resolution over the area of active star formation in a large nearby spiral galaxy in less than $\sim 10$--$20$~h, allowing for the prospect to survey a large set of $z = 0$ galaxies.

\item The ability, via integration with VLBI capabilities, to measure the proper motion of Andromeda's nuclear source at $\approx 0.1\mu$as~yr$^{-1}$ and to measure distance to a weak Galactic maser at $d=20$~kpc via parallax.

\end{enumerate}

\noindent As with the whole white paper, we intend these proposed projects to spur discussion in the broader community that will help sharpen the proposed ngVLA into a powerful instrument for all astronomers, as well as an ideal complement to other next-generation facilities like 30-m class optical/near-IR telescopes, LSST, any new IR surveyor missions, large new single dish facilities, and the Square Kilometer Array.

\section{Harnessing the Full Potential of cm-Wave Spectroscopic Imaging}

A next generation VLA has the prospect to revolutionize how we view gas in and around galaxies (including our own) by making detailed cm- and mm-wave spectroscopy a survey tool. If the ngVLA is constructed in a way that emphasizes surface brightness sensitivity and multi-line spectroscopy then every observation of a galaxy or a Milky Way cloud would yield detailed diagnostics of the excitation, density and chemical state of the gas, as well as the presence and strength of photon- and/or cosmic-ray-dominated regions (PDRs, CRDRs) and shocks. Although the last decades have seen enormous advances in our knowledge of the structure of the cold interstellar medium (ISM) in the Milky Way and other galaxies, most of these advances have come from studying crude tracers of interstellar gas mass that tell us little about the physical state of the gas --- tools like the {\sc Hi} 21-cm transition or the low-$J$ $^{12}$CO lines. The chemistry and physical state of the cold ISM over large scales remains largely {\em terra incognita}. A ngVLA promises to change this more than any other planned or existing telescope.

Cold, mostly molecular gas is the material out of which all stars form and makes up most of the ISM over the active (star forming) region of most massive galaxies. This molecular gas feeds supermassive black holes and experiences feedback from star formation and AGN onto the ISM. As such, it undergoes shocks, dissociation, and often represents a key component (by mass) of galactic outflows and fountains. This cold gas thus exists in a wide array of physical and chemical states. Understanding the interplay of gas, stars, star formation, and feedback requires understanding these chemical and physical states. 

Spectroscopy of cm- and mm-wave molecular transitions allows one to investigate these physical conditions in much the same way that optical spectroscopy probes the internal physics of ionized gas. This power has been repeatedly demonstrated studying compact, bright targets (e.g., galactic nuclei, distant merging galaxies, Galactic cores). This work shows how these faint transitions from rare (compared to CO and H$_2$) molecules offer powerful diagnostics of temperature, excitation, density, shocks, and ionization. Specific diagnostics accessible within the proposed frequency coverage of the ngVLA include:

\begin{itemize}

\item {\bf Dense Gas Tracers} such as HCN, HCO$^{+}$, HC$_{3}$N, HNC, CH$_2$O, and CS. These high dipole moment species require gas volume densities $\sim 10^{5}$~cm$^{-3}$ or larger to excite. They thus offer a direct window into gas volume density, often considered to be the main driver of the ability of gas to form stars.

\item {\bf Gas Excitation Tracers} such as CO and its isotopologues, NH$_{3}$, and H$_{2}$CO. These lines are excellent probes of kinetic temperature, particular when observed at high spatial resolution. The kinetic temperature, in turn, is vital to understand gas energetics and thermal pressures.

\item {\bf Shock Tracers} such as SiO, CH$_{3}$OH, and HNCO, species whose gas-phase abundance is sensitive to shocks (usually via their interaction with dust grains). Their detection indicates the presence of shocks and inter-comparisons between them provide insights into shock strength and the degree of photodissociation present.

\item {\bf Photon- / X-ray - / Cosmic Ray-Dominated Region Tracers} such as CN, C$_{2}$H, c-C$_{3}$H$_{2}$, CH, and HNC are catalyzed in the presence of ions (especially C+) and maintain significant abundances in regions where ionization sources are strong. Contrasting these with other species offers a diagnostic of the degree and type of ionization dominating a region and so, for example, to map out both the extent and influence of AGN and massive star formation on the ISM.

\item {\bf Complex Organics} are larger molecules that are created by more circuitous reaction pathways, including CH$_{2}$CO, HCOOH, CH$_{3}$CHO, NH$_{2}$CHO, CH$_{3}$C$_{2}$H, CH$_{2}$CH$_{2}$CN, CH$_{2}$CH$_{2}$OH and HCOOCH$_{3}$. Their presence and abundance offer key constraints on astrochemical modeling. Once understood, these models offer a  unique window into the density and temperature history of gas.

\end{itemize}

\noindent These spectroscopic probes offer an essential window into the physics of cold gas. However, their use has so far been limited in scope because in many environments of interest, cm- and mm-wave transitions beyond the bulk gas tracers are simply too faint to survey effectively with current instruments. Even the productive ammonia and formaldehyde mapping efforts using the current VLA and the first ``Band 3'' ($84$--$116$~GHz) results with ALMA demonstrate how time intensive cm- and mm-wave spectroscopic imaging remains. If spectroscopy of the cold universe is taken as a major driver for ngVLA design, then by concentrating a large amount of collecting area in the inner $\sim$ km of the array the ngVLA has the potential to open up this part of the spectrum, making it possible to deploy detailed spectroscopy of cold gas to study all types of galaxies across the history of the universe. 

\begin{figure}[!t]
\centering
\includegraphics[height=14cm]{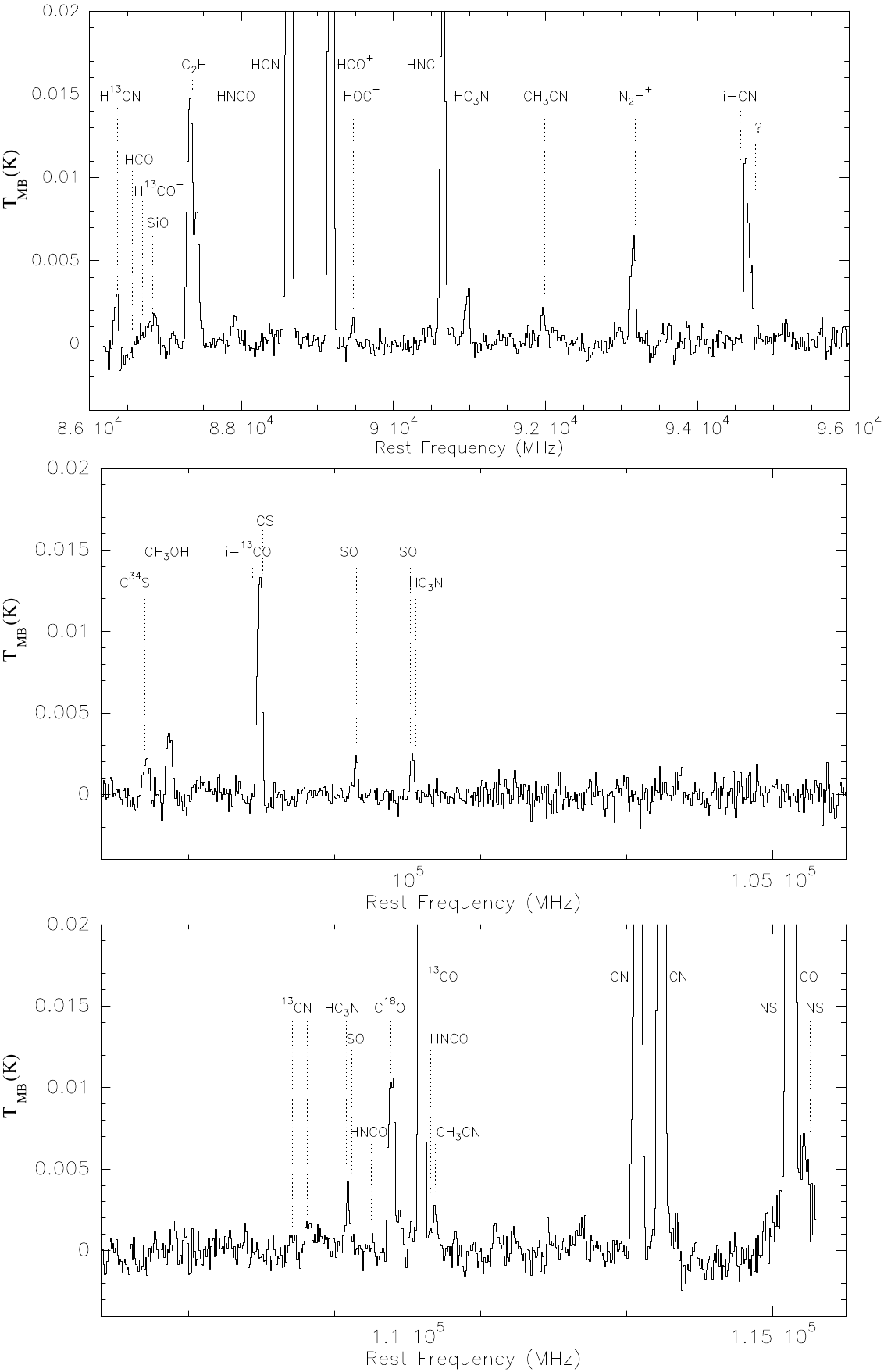}
\caption{{\small {\bf The rich 86-115~GHz spectrum.} From Aladro et al. (2013) -  this spectral scan of the nucleus of NGC 1068 shows the rich spectrum over the range 86-115~GHz. The next generation VLA could have the sensitivity to quickly image all of these transitions at $\approx 1\arcsec$ resolution across the whole area of a normal galaxy or Galactic star forming region. These lines trace shocks, gas density, excitation, ionization and the UV field, and offer the chance to harness the growing field of astrochemistry as a main tool to understand the physics of the interstellar medium, star formation, and galaxy evolution.}
\label{fig:n1068spec}}
\end{figure}

The same sensitivity gains that will open new transitions to survey science will make the ngVLA a phenomenal facility to pursue {\em high angular resolution imaging of bright transitions}. Because of the trade-off between angular resolution and sensitivity inherent in a fixed-collecting-area interferometer, high resolution imaging of thermal lines (like CO and {\sc Hi}) tends to not be limited by the availability of a few long baselines, but by the need for large amounts of collecting area at these baselines in order to achieve good surface brightness sensitivity. That is, for an array configuration plan optimized for surface brightness sensitivity (see below), a facility able to survey faint lines at $\sim 1\arcsec$ resolution will also be able to image brighter lines at $\sim 0.1\arcsec$ resolution.

If optimized for wide-area mapping of a suite of essential molecular gas tracers at $\sim 1''$ resolution (and higher for bright lines), the ngVLA would deliver our best window yet into the structure and physics of cold gas in galaxies. Major areas of improvement would include:

\begin{itemize}

\item {\bf Physical Conditions in the Cold ISM:} The microphysics of the cold ISM (temperature, pressure, density) are linked to the macrophysics of the gas (dynamics, galactic structure). In turn, these microphysics control its ability to form stars. Currently, our knowledge of the patterns of density, temperature, shocks, and pressure across galaxies is severely limited. What we do know tends to be focused on the warm, dense phase of the molecular gas, which is easier to observe but not representative of most of the gas mass. By probing these conditions in the cold gas that makes up most of the molecular ISM in normal galaxies, the ngVLA will be the ideal observational facility to change this. In local galaxies, it will resolve the changing physical conditions in the cold ISM and so measure how gas microphysics is affected by passage through spiral arms, spurs, bars, and other dynamical features.

\item {\bf An External View of ISM Structure at High Physical Resolution:} Just like the microphysical conditions, the structure of the cold ISM is clearly linked to both galactic environment and star formation. For example, massive clouds appear to preferentially form in spirals arms. In the Milky Way, star-forming dense cores ($\sim 1$~pc; Lada \& Lada 2003) appear to preferentially grow inside molecular clouds in spiral arms (Sawada et al. 2013). Within local molecular clouds star formation appears closely related to structures of dense filaments (width $\sim 0.1$~pc).  However, our ability to connect cold ISM structure to the wider galactic environment is severely limited by our restrictive point of view within the disk of our Galaxy and our inability to resolve the ISM in other galaxies. The ngVLA promises to change this by making it possible to image cloud sub-structure in galaxies out to $\sim 2$--$5$~Mpc (e.g., imaging CO emission at $0.1\arcsec$). This fraction-of-a-parsec imaging would enable the study of extragalactic clouds in a way currently restricted to Milky Way gas, but also to be cleanly correlated with galactic environment and star formation. Not only mapping, but metrics like the brightness distribution and the identification of filamentary structure would be accessible. Deploying high resolution Milky Way tools beyond our internal perspective will answer questions like: How does molecular gas evolve through different types of spiral arms and bars? How do clouds evolve between spiral arms? How important is ``triggered'' star formation? How is star formation suppressed in some molecule-rich environments, such as our own Galactic Center?

\item {\bf Challenging the Giant Molecular Cloud Paradigm:} Current studies of the molecular ISM in galaxies usually frame their analysis in terms of giant molecular clouds (GMCs) --- discrete, gravitationally bound, predominantly molecular structures. However, the validity of this paradigm is only weakly demonstrated. Are GMCs real entities or just one scale in a continuous turbulent power spectrum that appear discrete thanks to chemical and observational biases? Does the answer to this question depend on the type of galaxy or location within a galaxy? Testing the GMC paradigm requires high surface brightness sensitivity ($\sim 1$~K), high spatial and velocity resolution ($\sim 1$~pc and $\sim ~1$~km/s) observations to recover ISM structure across a large dynamic range. With such observations of a variety of galaxy types, one could measure the shape, dynamical state, and power spectrum of ISM structures across a wide range scale, making a strong test of the prevailing ``GMC paradigm.'' The ngVLA promise of high surface brightness sensitivity at high resolution are the key enabler here; even ALMA CO observations quickly hit the limit of surface brightness sensitivity at such high resolutions.

\item {\bf Differentiated Chemistry in Galaxies:} We know little about how molecular gas chemistry varies across whole galaxies. The ngVLA will map emission from simple and complex molecules as a function of location in a galaxy and --- as with physical conditions and gas structure --- allow us to connect chemistry to galaxy dynamics and galactic structure. There is still a huge amount to learn about the astrochemistry of normal cold gas and the utility of molecular emission to trace physical conditions.

\item {\bf Densities and Temperatures from Formaldehyde and Ammonia to 5 Mpc:} Formaldehyde, with transitions at 4.8 and 14.4 GHz, and ammonia, with transitions near 23 GHz, are two of the most powerful tools to robustly measure physical conditions in the molecular ISM. Specifically, observations of multiple cm-wave transitions of these two molecules offer direct, robust probes of density (formaldehyde) and temperature (ammonia) in dense, pre-cluster forming gas. Because they lie in the cm-wave regime, these molecules are well-suited to trace cold gas (unlike analogous sub-mm transitions, which preferentially trace excited gas). Currently these lines are too faint to be observed outside of the Milky Way or bright galaxy centers. With the sensitivity of the ngVLA, these key diagnostics could be applied to normal molecular gas in galaxies out to $\approx 5$~Mpc (the nearest few groups). This would provide a powerful view linking the formation of proto-clusters to galactic environment that has so far been out of reach for extragalactic studies.

\item {\bf Observing the Feeding and Feedback of Supermassive Black Holes (SMBHs):}  Based on the black hole-bulge relation, supermassive black holes are expected to strongly couple to their host galaxies and, indeed, feedback on the molecular gas is visible in the form of outflows and possibly suppressed star formation. With extremely high resolution spectral line mapping, the ngVLA would offer a powerful tool to observe the feeding of black holes via gas inflow, to resolve ``molecular tori'' in the nearest systems, and to probe the innermost AGN environment via absorption. It would also be a powerful tool to highlight the black hole's zone of influence on the surrounding gas by observing altered chemistry, X-Ray dominated region tracers, and increased turbulence and gas excitation. Similarly, the spectroscopic capabilities of the ngVLA will allow for the measurement of the detailed chemistry and physical conditions in the molecular outflows now observed regularly around SMBHs, including observations constraining the mass loss rate, the presence and strength of shocks and detailed observations of the outflow morphology. Sub-mm ALMA observations are already being utilized to explore many of these phenomena in the energetic regions close to black holes. The ngVLA, through its sensitivity to low-excitation gas, will allow exploration of these processes using an extended set of tracers (ones that focus on the cold, low excitation-medium), constraining the area of influence of the black hole on the surrounding galaxy and halo.

\item {\bf Magnetism Across the Molecular ISM:} Magnetic fields are known to be important to the molecular ISM: some filamentary structures are observed to lie perpendicular to magnetic field lines, and field strength is comparable to other terms in the energy budget for molecular clouds.  However, we still have only sparse measurements of the field strength and orientation in the molecular ISM, and our knowledge of the exact role of the magnetic field is consequently limited. What we do know comes from Zeeman-active species such as CN and OH, both of which are in the ngVLA band and the strength of field that can be probed is limited by telescope sensitivity. The proposed sensitivity of the ngVLA will enable Zeeman mapping that could be used to probe the field strength and orientation across wide areas in clouds. This kind of resolved magnetic field structure will be crucial to finally pinpointing how the field affects ISM evolution.

The ngVLA sensitivity would also open up mapping of the Goldreich-Kylafis effect, which causes rotational lines common of species such as CO and CS to show polarization at the few percent level. This would finally map the field strength and orientation in the bulk of gas in molecular clouds, spanning the range between the diffuse gas probed by OH Zeeman measurements to the dense gas sampled by CN Zeeman measurements. The Goldrich-Kylafis effect is maximized near the critical density of the lines in question, so that using a variety of lines in the ngVLA window one would be able to map the field strength through the molecular cloud.

\item {\bf New Views of Turbulence from Proper Motions of the ISM:}  Turbulence dominates the motions of the cold ISM and dictates the density and kinematic structure of star-forming clouds, but turbulent motions of gas in clouds are usually only observed along one dimension (the line of sight). Applying the exquisite sensitivity of the ngVLA to observe bright lines toward nearby clouds, the ngVLA will allow for proper motion studies of thermal emission from gas in turbulent ISM clouds. Although expensive in terms of observing time, such observations would, for the first time, observe the full three dimensional velocity structure of turbulence in the ISM. Such observations would directly answer a suite of questions about how turbulence leads to star formation: for example by distinguishing between compressive and solenoidal motions or by directly measuring the velocity power spectrum and its associated anisotropy.

\item {\bf The Organic Universe:}  Large molecules, many of potential pre-biotic importance, preferentially have transitions in the cm regime. Observing these large molecules constrains the degree of chemical complexity in the ISM and mapping their presence can tell us about their origin: Where do large molecules form? Organics likely are delivered to Earth via comets, asteroids and other similar solar system bodies, but did the larger molecules formed in situ in the Solar System or instead collected there from earlier formation in the ISM?  Does their abundance vary with location in the Galaxy, perhaps leading to ``chemical Galactic habitable zones'' that depend, e.g., on metallicity? The sensitivity and wide frequency coverage of the ngVLA will allow the exploration of the abundance and potential diagnostic power of complex organic molecules via their cm-wave transitions.

\item {\bf Tracing the Hidden Diffuse Molecular ISM:} The virtual invisibility of H$_{2}$ is one of the largest obstacles to studying the molecular ISM. Although CO and many other molecules can serve as good tracers of dense, well-shielded H$_2$, tracing low-density, less well shielded gas (``diffuse'') gas is more challenging. This regime is important: it may hold a large fraction of the total molecular gas mass (the ''hidden'' or CO-faint molecular gas) and is key to understanding the transition from atomic to molecular gas. In this regime the simple molecules CH and OH offer the prospect to selectively trace gas in this ``diffuse'' regime at the boundary of the atomic and molecular phase. Again, such observations are currently limited by sensitivity but it would be possible to survey whole galaxies for emission from diffuse molecular gas using the proposed ngVLA.

\end{itemize}

\noindent These examples show how, {\em if optimized for thermal line sensitivity}, a next-generation VLA could qualitatively change the kind of science done using microwave spectroscopy. Instead of surveys of bulk tracers and detailed spectroscopy of bright objects, such a facility would allow detailed spectroscopic surveys. Deployed across the field of low redshift galaxies, Milky Way ISM, and star formation studies this capability will be revolutionary. If we had such a machine {\em now} we would be in a position to straightforwardly understand the role of density in star formation, the evolutionary sequence of molecular clouds and cluster-forming cores, the universality of filamentary structure, the gas density and temperature distribution within the Milky Way and other galaxies, the physics of molecular outflows from galaxies, and the nature of the diffuse molecular ISM. The same power would render bulk gas tracer observations so routine that one could envision matching the $\sim 10^4$ galaxy sample sizes of current optical IFU surveys like MANGA or SAMI, and observing the evolution of gas fraction, dense gas content, and molecular outflows since $z \sim 1$ would be straightforward. In fifteen years the pressing questions will have evolved, but there is no doubt that the scientific leap will be similar: this is a frequency range rich in diagnostic capability where present generation instruments are starved for sensitivity. \\

\noindent {\bf Applications:} Given a ngVLA optimized for high sensitivity line work it will be possible to:

\begin{itemize}
\item Survey a large part of the area of active star formation in the inner part of the Milky Way at resolution $\sim 1\arcsec \approx 0.04$~pc and sensitivity to the entire range of physical processes granted by the rich cm- and mm- spectrum. At the distance of the Milky Way center and the star-forming Galactic Ring, this resolution is enough to resolve filamentary structure within molecular clouds and other aspects of detailed cloud structure. This kind of multi-tracer, high spatial resolution view is only currently achievable in targeted local clouds or over small areas. The simultaneous line survey speed and good resolution of the ngVLA would be an order of magnitude improvement over anything achievable by existing telescopes.

\item Survey a large sample of galaxies in tracers of shocks, gas density, excitation, ionization, and a host of other physical parameters at $\sim 1\arcsec \approx 50$~pc resolution. Such a survey would measure the life cycle of gas in galaxies, the impact of feedback on interstellar gas, how stellar nurseries couple back to their galactic environments, and the origin and evolution of molecular gas flowing out of galaxies. For the nearest ($\lesssim 20$~Mpc) galaxies, such a survey would yield a highly detailed spectrum of each individual molecular cloud in each galaxy, providing a truly ``top down'' view of star formation using diagnostic tools (faint mm- to cm-wave lines and continuum) only previously accessible to Milky Way studies. For more distant galaxies where Milky Way-like resolution is simply not achievable, this sophisticated spectroscopic view offers the best possible access to the small-scale physics of the ISM.

\item Use absorption against Galactic and extragalactic continuum sources across the sky (almost ``for free'' for some telescope designs) to assemble a detailed, unbiased probe of the extended molecular medium in the Milky Way, the dense areas around ultracompact {\sc Hii} regions, and (via absorption at the redshift of the source) the internal molecular medium around quasars and AGN.

\end{itemize}

\begin{figure}[!t]
\centering
\begin{minipage}{0.40\textwidth}
\centering
\includegraphics[height=13cm]{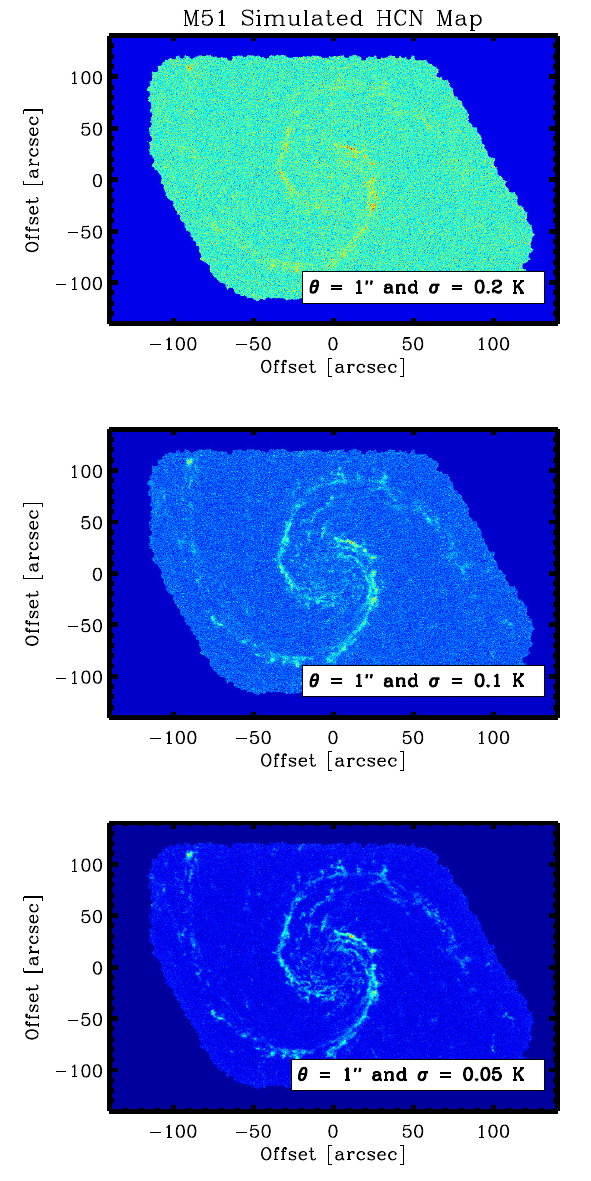}
\end{minipage}\hfill
\begin{minipage}{0.475\textwidth}
\centering
\caption{\small {\bf The Need for High Brightness Sensitivity.} Here we show a simulated map of HCN at $1\arcsec$ in M51 based on scaling the PAWS CO survey by Schinnerer et al. (2013) by a typical HCN-to-CO line ratio (1/20). We show the structure of the galaxy at a series of surface brightness sensitivities --- $0.2$~K, $0.1$~K, $0.05$~K. M51 is among the brightest normal nearby galaxies and HCN among the brightest diagnostic transitions other than the CO lines, but it is already clear that good surface brightness sensitivity, much better than $< 0.1$~K must be achieved quickly to make sophisticated microwave spectroscopy a regular survey tool. With the proper distribution of collecting area, the ngVLA could achieve high surface brightness sensitivity and coverage of a huge part of the mm- and cm-wave spectrum simultaneously, advancing the field significantly beyond ALMA's already impressive capabilities. For sophisticated mm- and cm-wave spectroscopy to be major science driver the telescope will need to be designed in a way that dramatically exceeds ALMA's effective collecting area on baselines $\sim 1$~km. 
\label{fig:hcn}}
\end{minipage}
\end{figure}

\noindent {\bf Technical considerations for the ngVLA to be revolutionary:} The brightest set of interesting diagnostic lines beyond CO (HCN, HCO$^+$, CS, CN, etc.) are $\sim 10$--$30$ times fainter than CO in a typical star-forming galaxy. Following the radiometer equation, a line $N$ times fainter than a bulk gas tracer, say CO, requires $N^2$ times the integration time to reach matched signal to noise. This means that to achieve detailed multi-line spectroscopy of quality comparable to current CO observations would require $\sim 100$ to $1,000$ times the time investment required to make the CO maps on the same instrument.

For thermal, optically thick lines like CO in the cold ISM, a fiducial brightness of $\sim 1$~K for a marginally resolved target is reasonable (a typical cold cloud is $\sim 10$--$20$~K, so this allows for substantial subtructure). Before ALMA, the best telescopes in the world required $\sim 200$\,hr to map CO at $\sim 1\arcsec \times 10$~km~s$^{-1}$ resolution and a few times 0.1~K sensitivity across a nearby galaxy. Achieving similar quality maps of the full suite of diagnostic lines would require an impossible 20,000--200,000 hours on the previous generation of forefront facilities.  In bright regions, and through heroic efforts at lower resolution, these diagnostics are now being explored, but their use as a general purpose tools will remained limited by these simple sensitivity concerns.

Figure \ref{fig:hcn} demonstrates this concretely. The Figure shows an arcsecond resolution CO map of M51, taken from the PAWS survey, scaled by an appropriate line ratio to resemble an arcsecond HCN map of the same galaxy. The Figure then illustrates the peak intensity of this HCN map observed at an arcsecond resolution with rms noise $0.2$, $0.1$, and $0.05$~K. M51 is among the brightest disk galaxies in the sky but clearly a fraction of a Kelvin sensitivity is needed to study the galaxy in even the brightest diagnostic lines beyond CO. Creating the CO map itself with the Plateau de Bure interferometer required $\sim 200$\,hr and the achieved noise was $\sim 0.4$~K per $5$~km~s$^{-1}$ channel; improving this to $0.1$~K or $0.05$~K would require thousands of hours of integration time. 

{\em Relation to ALMA:} ALMA improves the situation dramatically but in the $\nu = 70$--$115$~GHz range the science described here is still incredibly time consuming. In the sub-mm  regime, ALMA shatters previous capabilities and is a stunning tool to study dust emission, excited gas, and active regions. Though still revolutionary at $\nu \sim 70$--$115$~GHz, ALMA in its current (nearly final) form would still require $\sim 7$ hours on source per field to produce a $1\arcsec$ resolution CO map of a nearby galaxy at the $\sigma = 0.05$~K, $\Delta {\rm v} = 5$~km~s$^{-1}$ level. Surveying any appreciable area to this depth and pushing fainter (recall that M51 is among the brightest plausible targets and HCN among the brightest plausible transitions) increases the time requirements rapidly. So one can reasonably expect ALMA to revolutionize this area, but the intrinsic faintness of the lines under discussion mean that deploying sophisticated microwave spectroscopy as a survey tool will need to wait for a next generation facility, possibly the ngVLA, or focus on the sub-mm wavelengths, where ALMA's noise in brightness temperature units is much better thanks to the Rayleigh Jeans scaling (and for fixed velocity resolution, a larger $\Delta \nu$).

This sub-mm regime is key, but as we have emphasized above, {\it the $70$ to $115$~GHz range is uniquely important to understand the cold ISM}. This frequency region contains a high density of powerful diagnostic lines (described above). Because these have low excitation requirements ($h \nu / k_B \sim 4$~K), they are among the most generally useful and widespread (in terms of emission) transitions for each of these molecules. Indeed, for the very interesting cold ($\sim 10$~K) component of the star-forming ISM these lines are almost the  only option to explore the physical conditions. 

{\em Requirement for a revolutionary spectroscopy machine:} Following this argument the first-order technical requirement for the ngVLA to be a revolutionary spectroscopy machine are straightforward: it must vastly outstrip the spectroscopic capabilities of ALMA in the range $20$--$115$~GHz. Practically this means that the {\bf surface brightness sensitivity} of the array must be much higher than that of ALMA for resolutions $0.1$--$1\arcsec$ given a fixed amount of observing time. Going forward, this {\em must} be the case, otherwise, given the likely construction date of the ngVLA and plausible upgrades to ALMA's receiver capabilities (including extensions to lower frequency and perhaps wider bandwidth), much of the exciting spectroscopic ngVLA science in this area could possibly be ``done'' by the date of its construction.

Fortunately, the prospects for an ngVLA that opens new parameter space via its surface brightness sensitivity --- and one that specifically enables the diagnostics discussed here ---  are excellent. The collecting area under discussion (see the introduction) is large compared to that of ALMA (indeed the physical collecting area of the VLA already exceeds that of ALMA). The key to enabling the science discussed here (and indeed much of the science focused on studying the ISM at all redshifts) is that {\em a large amount this collecting area must be situated at relatively compact baselines, $\approx 1$~km.} Note that this does differ from the notional ngVLA discussed in the cover letter, which exceeds ALMA's effective area at overlapping frequencies, but places only $\sim 20\%$ of this area at short baselines. This notional ngVLA in fact only roughly matches ALMA's capabilities at $1\arcsec$ resolution. Although it exceeds the capabilities of ALMA at higher resolutions ($\sim 0.1\arcsec$) it does not do so yet by a large factor. Given the surface brightness sensitivity at these frequencies are best suited to highly resolved objects or bright lines, one may also wonder whether the comparison to the brightness temperature sensitivity of the upper ALMA bands is more relevant.

The second-order considerations to optimize the ngVLA for spectroscopic science are also straightforward. First, gas in galaxies has a complex, multi-scale distribution and the capability to make full-fidelity, full-flux-recovery images should be included as a baseline part of the design. ALMA accomplishes this by mixing dish sizes, but an equally good strategy might be to pair with a large single dish telescope. Second, the spectroscopic science discussed here relies heavily on combinations of lines and combinations of lines and continuum observations. To that end, a wide instantaneous bandwidth that still offers the spectral resolution needed to resolve individual transitions is key. Covering many science goals in a single observation multiplicatively increases the power of the telescope. Finally, key transitions span all the way from $1$ to $116$~GHz and the science described here would benefit immensely from full coverage of this frequecy range.

\section{Resolving Hidden Power Sources --- Massive Star Formation and the Feeding of Supermassive Black Holes}

\noindent The ability to image thermal emission at high spatial resolution is also revolutionary. From forming stars to the centers of galaxies, to the immediate environs of black holes, a large cm- and mm-wave interferometer would offer the ability to resolve the motions and structure of the ionized and cold interstellar medium through heavy extinction and with unparalleled sharpness.\\

{\bf Applications to High Mass Star Formation} \\

{\bf Finding and Resolving Massive Pre-Stellar Cores:} Sensitive observations at millimeter to centimeter wavelengths at a range of angular scales are essential to understand the formation of massive protostars and protoclusters.  On the largest scales, a network of infrared dark clouds (IRDCs) span the Milky Way, and many of them contain multiple sites of active massive star formation spread over several arc-minutes. In recent years, the earliest phase of candidate massive pre-stellar cores have been identified within IRDCs in the form of low temperature ($<20$K), high column density, compact ($1000-4000$\,AU) objects prior to collapse of the first protostars. Thanks to these cold temperatures, kinematics of these objects are best studied in the low-excitation molecular transitions in the $3-4$\,mm band, including using deuterated species whose production is enhanced in cold interstellar clouds.

A next-generation VLA, such as the one discussed here, will enable such kinematic investigations targeting a large population of pre-stellar cores in a wide set of transitions. To achieve this, observations with high spatial resolution  (0\farcs25$= 500-2000$\,AU at $2-8$\,kpc) but still high enough surface brightness sensitivity to image thermal, low excitation lines are required. Although ALMA can reach these resolutions in the $\lambda = 3$--$4$~mm window now, its total collecting area is insufficient to image a large set of such low brightness temperature sources in detail in a reasonable amount of time. The ngVLA would allow wide bandwidth (covering many lines), high brightness sensitivity, high spatial resolution surveys of a large population of these massive pre-stellar cores, capturing their chemical and kinematic evolution and providing a strong anchor for theoretical models.

\begin{figure}[!t]
\begin{center}
\includegraphics[width=0.8\textwidth]{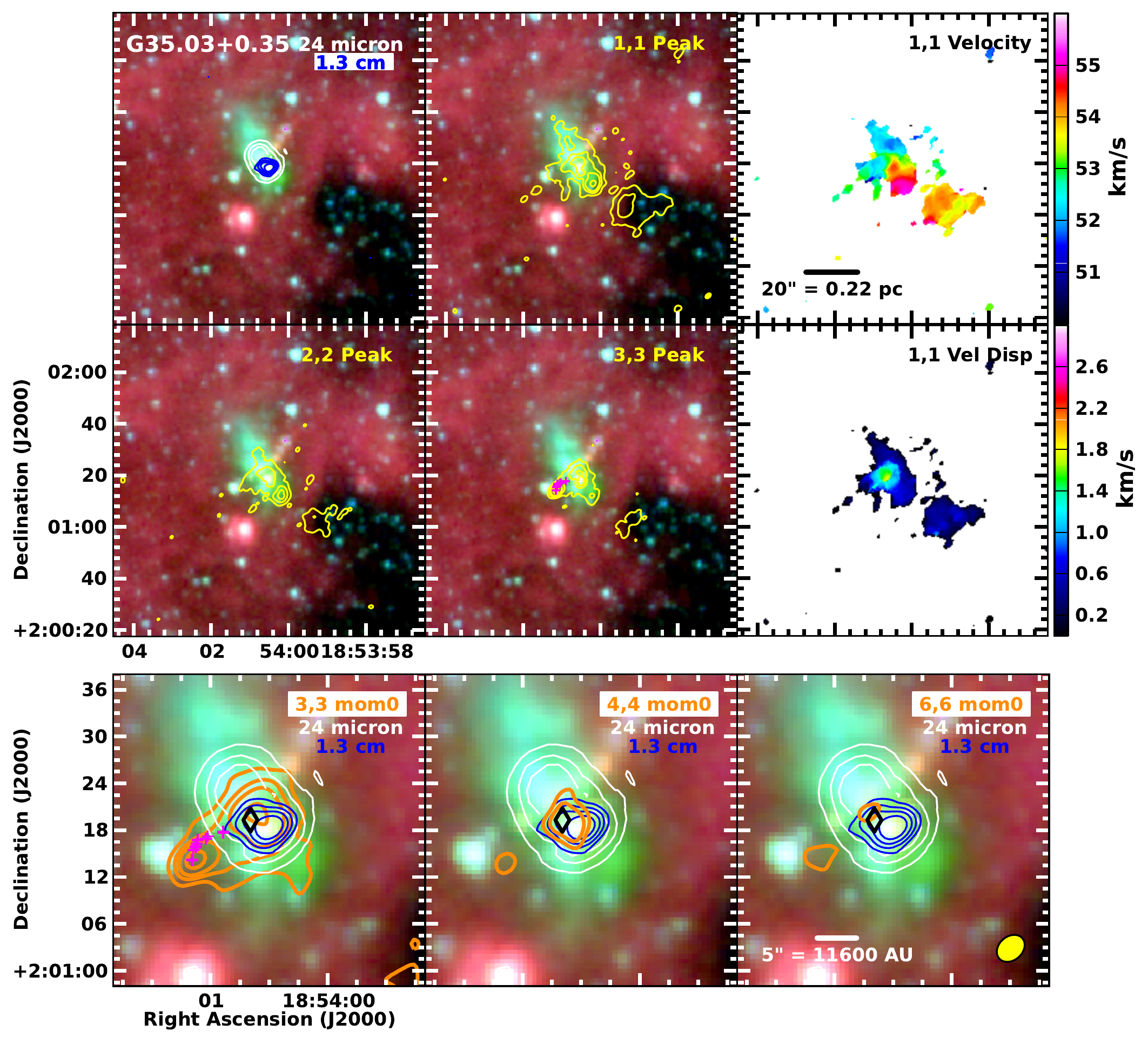}
\end{center}
\caption{\small JVLA images of the NH$_3$ transitions and continuum emission from the massive protocluster G35.03$+$0.35 using the D-configuration. While providing copious diagnostic information like temperature and kinematics, it is time-consuming to push these types of observations to higher angular resolution due to surface brightness sensitivity limitations.  It will take a ngVLA to make such investigations routine and study lower-mass protoclusters.  }
\label{fig:G35}
\end{figure}

\vspace{6pt}
\noindent{} {\bf Protoclusters and accretion:} The collapse of massive prestellar cores leads to massive protostellar clusters (protoclusters), which consist of multiple deeply-embedded dust cores, some with hypercompact or ultracompact H{\sc ii} regions, clustered on scales of a few thousand AU.  The detailed study of the accretion structures around individual massive protostars requires tracers of warm ($>$100\,K), high density gas, whose high-$J$ transitions are typically associated with the sub-mm wavelengths accessible with ALMA.  However, when studying the central massive protostars and their accretion structures, {\it long millimeter wavelengths ($7-13$\,mm) are essential for penetrating the high continuum opacities from dust grains} that will hamper shorter wavelength spectral observations with ALMA.  

The transitions of thermal ammonia and methanol in the $\lambda = 10-13$\,mm range will be key probes of accretion kinematics, in addition to the many maser transitions of water, ammonia and methanol.  At these wavelengths, the current VLA maximum baselines of 36\,km are sufficient to reach scales of $0.075\arcsec$ ($150-600$\,AU at $2-8$\,kpc), which is essential to resolve accretion onto a single massive protostar vs. a proto-binary. However, the current collecting area of the VLA on these baselines is insufficient to detect cool thermal emission associated with these transitions (the current VLA struggles with these transitions beyond its ``C'' configuration). The large increase in collecting area promised by the ngVLA is essential to utilize these longer baselines for thermal emission. For example, the notional ngVLA described in the associated cover letter would be able to reach an rms noise of 20\,K per 1.5\,km/s channels in 12\,hr, allowing one to resolve and distinguish between single/binary protostars. Although ALMA could theoretically reach these size scales at the upper end of ALMA Band 1, it can only do so in the continuum due to the smaller collecting area, and even then only out to the distance of the nearest high mass star forming region (Orion) and so cannot access the bulk of Milky Way massive star formation.

\vspace{6pt}
\noindent{} {\bf Ionized jets and hypercompact H{\sc ii} regions:} The cm continuum emission from massive protostars traces bipolar ionized jets, such as those seen in the Cepheus\,A cluster, which along with molecular outflows provide feedback into the surrounding core.  As they approach the main sequence, these stars power hypercompact H{\sc ii} regions which eventually expand and potentially impact the formation process of neighboring protostars. These ionized phenomena are best observed at wavelengths of $\lambda = 1-7$\,cm. In order to achieve continuum images at the longest wavelength in this range that match the $0.075\arcsec$ resolution molecular gas images at 1.3\,cm, the ngVLA will require baselines of 200\,km. In addition, because the ionized gas is of higher temperature ($>$5000\,K), the 200\,km baselines can also be exploited at the shorter wavelengths for continuum. For example, at 1.5\,cm, a beam of $0.02\arcsec$ can be achieved, providing $40-160$\,AU resolution at $2-8$\,kpc distance. Such images will pinpoint the jet launching point and/or measure the size of the hypercompact H{\sc ii} region, leading to accurate models of its density, pressure, and the required infall rate to quench its expansion.  Even with such a small beam, a more than adequate rms of 50\,K (1.6\,$\mu$Jy/bm) can be reached in only 5 minutes to resolve such compact regions. 

\vspace{6pt}
\noindent{} {\bf A complete census of the protocluster population:} Finally, and perhaps most importantly, the ngVLA will be able to detect the population of low-mass protostars and pre-main sequence stars forming in the immediate vicinity of massive protostars. This will make possible the first complete censuses of protostars in forming protoclusters. In addition to dust emission from circumstellar disks around low-mass protostars, which will be visible at $3-8$\,mm, the more evolved pre-main sequence stars often exhibit chromospheric activity detectable in the cm range. For example, several dozen stars have been detected in M\,17 (at 2\,kpc) with fluxes down to $\sim 0.1$\,mJy and in Cepheus\,A (at 700\,pc) at levels down to $\sim 0.3$\,mJy. These fluxes scale to $\sim 2$\,$\mu$Jy at 8\,kpc, within the sensitivity range of the ngVLA and demonstrating that the telescope as conceived above would have the ability to detect the active pre-main sequence stars in a protocluster as far away as the Galactic center in only a few hours.

\vspace{12pt}
\noindent
{\bf A High-Resolution Kinematic View of Nearby Galaxy Nuclei}\\

\vspace{-6pt}
\noindent{}
Species which give rise to abundant maser emission are useful for high-resolution studies of gas kinematics (and their implications for star formation), given the non-thermal brightnesses of these lines. A particularly useful example of this may be 36.2\,GHz collisionally-excited (Class I) methanol masers, which would be uniquely accessible to the ngVLA. Recent studies have shown (maser) emission from this line to be extremely abundant in clouds in the center of the Milky Way, and have detected it for the first time in an extragalactic source. Using the 36.2 GHz methanol maser line as a tracer, it could be possible to trace sub-pc structures (0.2-0.3\,pc) in individual clouds in galactic nuclei out to Maffei2 / IC342 / M82 (d $\sim$ 3-4\,Mpc), given an rms of $\sim$120 K at 180\,km baselines, with 10$\times$ the effective collecting area of the current VLA. For a single Galactic center cloud observed at comparable spatial resolutions with the VLA, $\sim$50 masers can be detected (5$\sigma$) with brightnesses 600\,K.  Thus far no other collisionally-excited methanol maser lines (e.g, those at 44, 84, or 96\,GHz, which will be accessible to ALMA) have been shown to be similarly abundant in any Galactic sources, though the 84\,GHz line has been detected as a megamaser in NGC\,1068.  

\begin{figure}
\begin{center}
\includegraphics[width=0.8\textwidth]{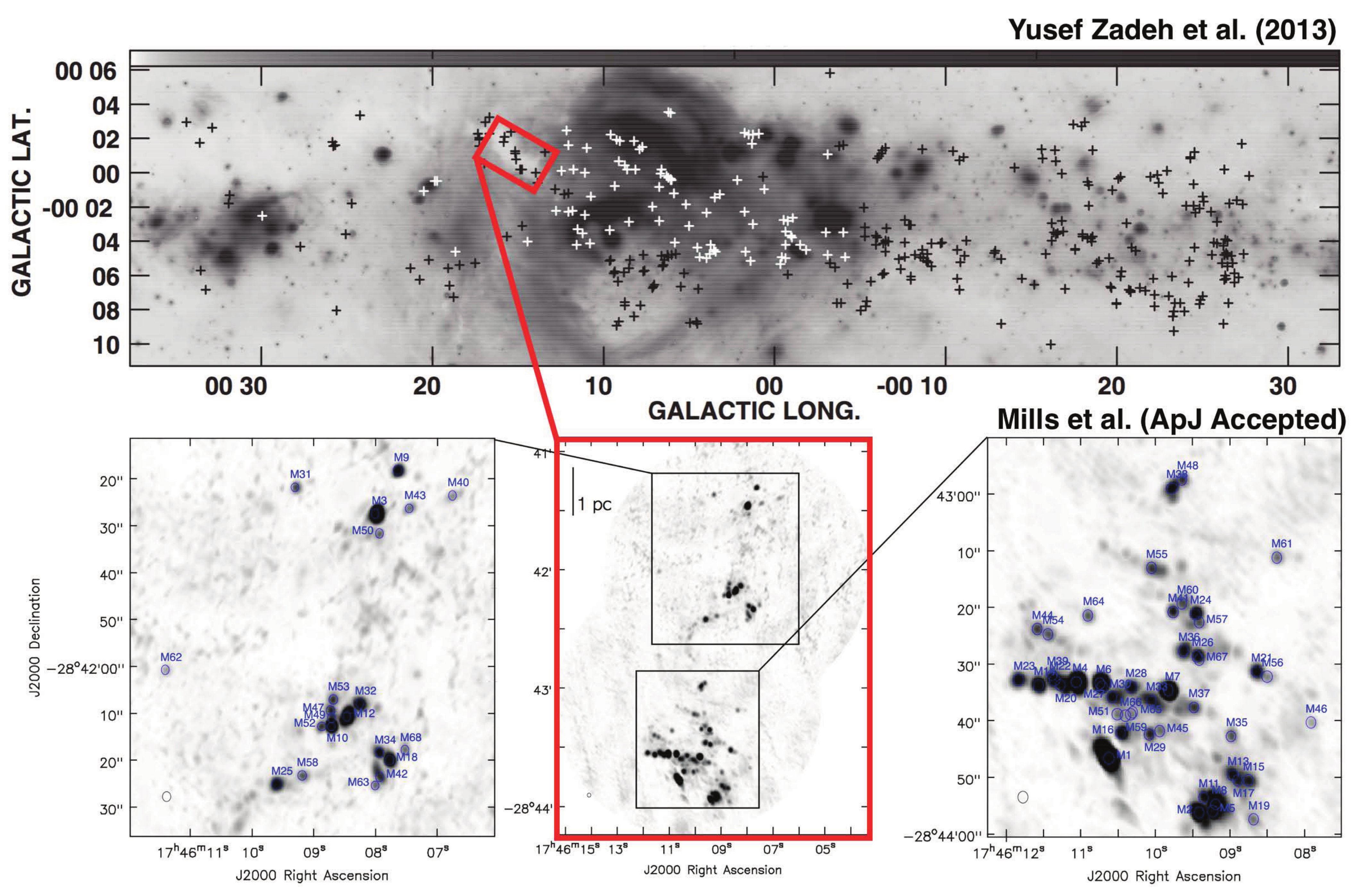}
\end{center}
\caption{\small
Top: More than 350 36.2 GHz methanol emission sources have been identified in the central 150\,pc of the Milky Way. Bottom: For one of these clouds, in which a lower spectral-resolution study of found 8 methanol sources, 0.1\,pc and 1\,km/s resolution VLA observations reveal nearly an order of magnitude more sources: 68 masers, all of which have brightness temperatures $>$400\,K. 
}
\label{fig:33GHz_masers}
\end{figure}

Given the observed preponderance of these masers in Galactic center clouds and environments where they are similarly common, these masers would allow for the kinematics and morphologies of individual clouds to be mapped on sub-pc or ``clump'' scales, potentially revealing filaments or other substructure (see Figure \ref{fig:33GHz_masers}). More quantitatively, the widths of these lines can be measured-- these are expected to be intrinsically sub-thermally narrow, but are observed in practice to be much broader, likely due to to emission from multiple masers. This would allow for measurements of the turbulent spectrum of the gas down to sub-pc scales, probing spatial variations of turbulence across these galaxies (or their centers). Observed variations in the properties of turbulence in the variety of extragalactic environments accessible with the ngVLA could yield important insight on the resulting star formation properties as a function of these environments, as theory and observations suggest that the initial conditions for star formation (e.g., the distribution of gas volume and observed column densities) are set by supersonic turbulence.  

Other species (e.g., methanol at 36.2, 37.1 and 84 GHz) have also been observed to exhibit megamaser emission in the centers of nearby galaxies. Surveys for such sources are thus far incomplete, so that their incidence in the local universe is not well constrained.  As maser emission is generally limited to a narrow range of physical conditions, variations in the in the chemical and physical conditions of Galactic nuclei can give rise to entirely different species and varieties of maser emission, e.g. OH or H$_2$O (mega)masers.  A next generation VLA would offer unprecedented sensitivity at radio frequencies to search for and characterize a larger range of potentially megamasing transitions (e.g., CH$_3$OH 24, 36, 44, 84, 96\,GHz) as well as more highly excited OH transitions observed to be masers in our Galaxy (at multiple frequencies between $\approx$1.6 and 13.4\,GHz). There thus remains a rich discovery space for the detection and characterization of additional megamasing transitions which will allow for a larger sample of sources for which accurate distances and black hole masses can be determined.

\vspace{12pt}
\noindent
{\bf Probing Black Hole Accretion around \sgra}\\

\vspace{-6pt}
\noindent{}

The H{\sc i} recombination line emission arising from the ionized gas accreting to \sgra and other nearby galaxies with super massive black holes will be an exciting possibility opened up by ngVLA. The sensitivity and angular resolution, combined with the avoidance of dust extinction plaguing optical studies of emission lines will be unique contribution of ngVLA. Detection of the H{\sc i} recombination line emission will open up an entirely new avenue to observe the central black hole environments -- enabling measurement of gas kinematics and  modeling of the accretion process by which the black hole is fueled.  This technique is unique  for studying the accretion zone. X-ray emission cannot probe $10^4$\,K gas (which is the likely accretion reservoir), and radio free-free emission is can be overwhelmed by the strong synchrotron emission. The recombination line emission also provides kinematic information vital to understanding the gas dynamics (spherical infall versus a rotating accretion disk), and to estimate the radius of the emission.

The mm recombination lines of H{\sc i} provide a {\bf reliable probe of dense, ionized gas in high extinction regions}. Recent theoretical work has derived the line emissivities as a function of principle quantum number, electron density ($n_e$) and temperature ($T_e$) for lines down to 200 GHz and these can be extended as a guide to emissivities at $>$ 50\,GHz, yielding emissivities in the range $10^{-32 \rightarrow -31}~n_e n_p ~ergs ~cm^{-3} ~sec^{-1}$. A critical difference between these high frequency lines and the longer cm-wave recombination lines results from the fact that they originate from lower {\it n} quantum levels where the spontaneous decay rates are much higher.  The recombination lines at high frequency ($>$ 50 GHz) have little or no population inversions in the levels, implying that the intensities can reliably be translated into source emission measures.  This makes them an excellent probe of {\bf highly obscured, high density ionized gas as must exist near AGN}. 

The radio source \sgra ~is identified with the central massive black hole in our Galactic center. The black hole mass, determined from the proper motions of closely orbiting stars, is $M = 3.7 \times 10^6~\msun $. The proximity of \sgra ~to Earth makes it our best candidate for detailed observations of the processes associated with buildup of massive black holes and their accretion. At present, the \sgra ~accretion rate is probably extraordinarily  small ($\mdot \sim 10^{-6}$~\msun~yr$^{-1}$ -- note that none of these are direct measurements). However, its accretion rate, averaged over the age of the Galaxy, must have been $\mdot \sim 4\times10^{-4}$~\msun~yr$^{-1}$ to build up its present mass.  \sgra ~is clearly the optimal source for understanding low accretion modes of AGN fueling yet to date we have lacked an adequate probe of the ionized gas.

\section{Unleashing the Diagnostic Power of the Full Radio Spectral Energy Distribution}

\noindent
Radio continuum observations have proven to be a workhorse in our understanding of the star formation process (i.e., stellar birth and death) from galaxies both in the nearby universe and out to the highest redshifts, albeit by typically relying on and having to interpret a measurement form a single frequency. A next-generation VLA would revolutionize our understanding of what powers the radio continuum emission in and around galaxies by enabling the routine construction of $\sim 1 - 100$\,GHz radio spectral maps, rather than niche observations for a few individual galaxies. Each observation will provide enough sensitivity and spectral coverage to robustly decompose and accurately quantify the individual energetic components powering the radio continuum, thus providing unique information on the non-thermal plasma, ionized gas, and cold dust content in the disks and halos of galaxies.  

\vspace{6pt}
\noindent
{\bf Characterizing the Energetic Processes Powering the Radio Continuum Emission within Nearby Galaxies:}\\

\vspace{-6pt}
\noindent{}
The microwave spectrum of galaxies covering $\sim 1 - 100$\,GHz is powered by mix of physical emission processes, each providing independent information on the star formation and ISM properties of galaxies. These processes include non-thermal synchrotron, free-free (thermal bremsstrahlung), anomalous microwave (e.g., spinning dust), and thermal dust emission. There are two main challenges to utilize this emission to study the physics of galaxies: the faintness of the emission and the challenge disentangling different emission mechanisms. The proposed ngVLA will address both issues and so promises a major step forward using continuum emission to study the physics of galaxies. Its large bandwidth makes it possible to disentangle the different emission mechanisms by observing a continuous large part of the radio spectral energy distribution, dealing with the main uncertainty for multi-frequency radio studies. Meanwhile the large collecting area and bandwidth of the ngVLA will allow detection of emission in normal galaxies, which has been too faint in the $\sim30-100$\,GHz frequency range to map widely in the general ISM of nearby galaxies using current facilities. This new facility would open up this frequency window to investigate these distinct physical processes across large, heterogeneous samples of nearby galaxies for the first time. 

\begin{figure}[!t]
\begin{center}
\includegraphics[width=0.8\textwidth]{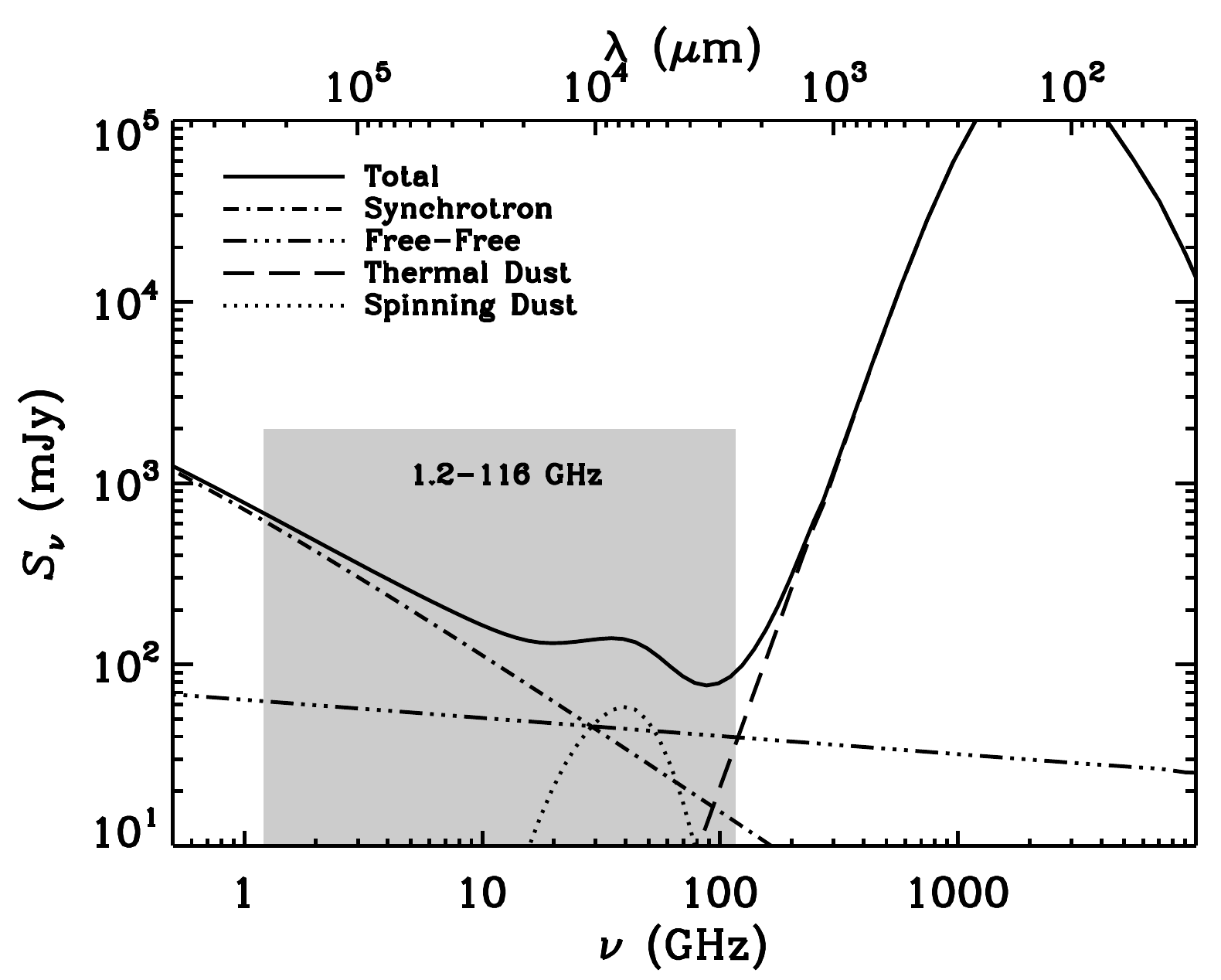}
\end{center}
\caption{\small A model spectrum illustrating the various emission processes (non-thermal synchrotron, free-free, spinning dust, thermal dust) that contribute to the observed microwave frequency range to be covered by the ngVLA. Only in the proposed ngVLA frequency range ($1.2-116$\,GHz, highlighted) do all major continuum emission mechanisms contribute at similar levels, making this range uniquely well-suited to next-generation continuum studies.  }
\label{fig:radspec}
\end{figure}

\begin{itemize}
\item {\bf Non-Thermal Synchrotron Emission:} 
At $\sim$GHz frequencies, radio emission from galaxies is dominated by non-thermal synchrotron emission resulting, indirectly, from star formation. Stars more massive than $\sim$8\,$M_{\odot}$ end their lives as core-collapse supernovae, whose remnants are thought to be the primary accelerators of cosmic-ray (CR) electrons, giving rise to the diffuse synchrotron emission observed from star-forming galaxies. Thus, the synchrotron emission observed from galaxies provides a direct probe of the relativistic (magnetic field + CRs) component of the ISM. As illustrated in Figure \ref{fig:radspec}, the synchrotron component has a steep spectral index, typically scaling as $S_{\nu} \propto \nu^{-0.85}$. By covering a frequency range spanning $1.2 - 116$\,GHz, the ngVLA will be sensitive to CR electrons spanning an order of magnitude in energy (i.e., $\sim1-30$\,GeV), including the population of CRs thought to be responsible for driving the dynamically important CR pressure term in galaxies.

\item {\bf Free-Free Emission:}  The same massive stars whose supernovae are directly tied to the production of synchrotron emission in star-forming galaxy disks are also responsible for the creation of H{\sc ii} regions. The ionized gas produces free-free emission which is directly proportional to the production rate of ionizing (Lyman continuum) photons and optically-thin at radio frequencies. In contrast to, e.g., optical recombination line emission, no hard-to-estimate attenuation term is required to link the free-free emission to ionizing photon rates. Unlike the non-thermal synchrotron emission, free-free emission has a relatively flat spectral index, scaling as $S_{\nu} \propto \nu^{-0.1}$. Globally, the free-free emission begins to dominate the radio emission once beyond $\sim$30\,GHz, exactly the frequency range that the ngVLA will be delivering more than an order of magnitude improvement compared to any current of planned facility.  

\item {\bf Thermal Dust Emission:}  
At frequencies $\gtrsim$100\,GHz, (cold) thermal dust emission on the Rayleigh-Jeans portion of the galaxy far-infrared/sub-millimeter spectral energy distribution can begin to take over as the dominant emission component for regions within normal star-forming galaxies. This in turn provides a secure handle on the cold dust content in galaxies, which dominates the total dust mass. For a fixed gas-to-dust ratio, this total dust mass can be used to infer a total ISM mass. Given the large instantaneous bandwidth afforded by the ngVLA, and more than an order of magnitude increase in mapping speed at 100\,GHz compared to ALMA, such an observations will simultaneously provide access to the $J=1\rightarrow0$ line of CO revealing the molecular gas fraction for entire disks of nearby galaxies. Alternatively, combining H{\sc i} observations (also available to the ngVLA) with $J=1\rightarrow0$ CO maps, one can instead use the thermal dust emission to measure the spatially varying gas-to-dust ratio directly.  

\item {\bf Anomalous Microwave Emission:}  In addition to the standard Galactic foreground components (free-free, synchrotron, and thermal dust emission), an unknown component has been found to dominate over these at microwave frequencies between $\sim 10-90$\,GHz, and is seemingly correlated with 100\,$\mu$m thermal dust emission. Cosmic microwave background (CMB) experiments were the first to discover the presence of this anomalous dust-correlated emission, whose origin still remains unknown. Its presence as a foreground still hampers studies as the accurate separation of Galactic foreground emission in CMB experiments remains a major challenge in observational cosmology. At present, the most widely accepted explanation for the anomalous emission is the spinning dust model in which rapidly rotating very small grains, having a nonzero electric dipole moment, produce the observed microwave emission. The increased sensitivity and mapping speed of the ngVLA will allow for an unprecedented investigation into the origin and prominence of this emission component both within our own galaxy and others, ultimately helping to improve upon the precision of future CMB experiments.    

\end{itemize}

\begin{figure}[!t]
\begin{center}
\includegraphics[width=12cm]{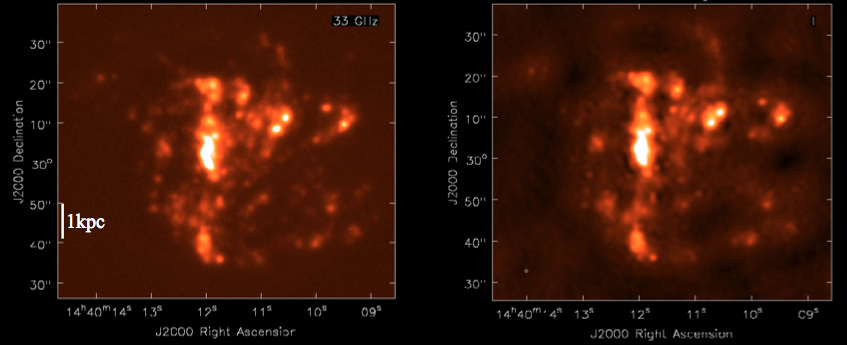}
\end{center}
\caption{\small Left: a model for the thermal Free-Free emission at 33\,GHz from NGC\,5713; $d \approx 21.4\,\mathrm{Mpc}, \mathrm{SFR} \approx 4\,\mathrm{M_{\odot}\,yr}^{-1}$). The model was estimated from H$\alpha$ imaging at a native resolution of $\approx$2\arcsec. The peak brightness temperature is 150\,mK, and the fainter knots are about 1\,mK. Right: The corresponding ngVLA image for a 10\,hr integration, with a bandwidth of 20\,GHz, centered at 30\,GHz and restored with a 1\arcsec~ beam.  The rms is 0.5\,$\mu$Jy\,bm$^{-1}$ (0.7\,mK), which is the equivalent of detecting free-free emission from H{\sc ii} regions comprised of as little as $\approx$2 O7.5 main sequence stars at the distance of NGC\,5713.  
\label{fig:n5713}}
\end{figure}

\noindent Clearly, almost all continuum science applications in the range 1--100~GHz would benefit immensely from constructions of the ngVLA. Simply taking the example of mapping a nearby galaxy in its free-free emission at 33\,GHz illustrates how the ngVLA is truly a transformational telescope for the entire astronomical community. Much of what we know about star formation in nearby galaxies on 1\arcsec~ scales is driven by observations of the Hydrogen recombination line H$\alpha$. However, the interpretation of H$\alpha$ narrow-band imaging is severely complicated by the presence of contaminating nearby [N{\sc ii}] emission as well as internal dust extinction, both of which vary spatially by large amounts within galaxy disks. Radio free-free emission maps at $\gtrsim$30\,GHz provide a direct measure of the ionizing photon rate associated with massive stars without having to make simplifying assumptions about such effects, and thus will yield robust measures for the star formation activity within nearby galaxies disks on $\sim$1\arcsec~ (i.e., $\lesssim$100\,pc) scales. Such maps will ultimately replace existing H$\alpha$ maps as the workhorse for studying the distributed star formation on $\lesssim$100\,pc scales in nearby galaxies.  

\begin{figure}[!t]
\begin{center}
\includegraphics[width=\textwidth]{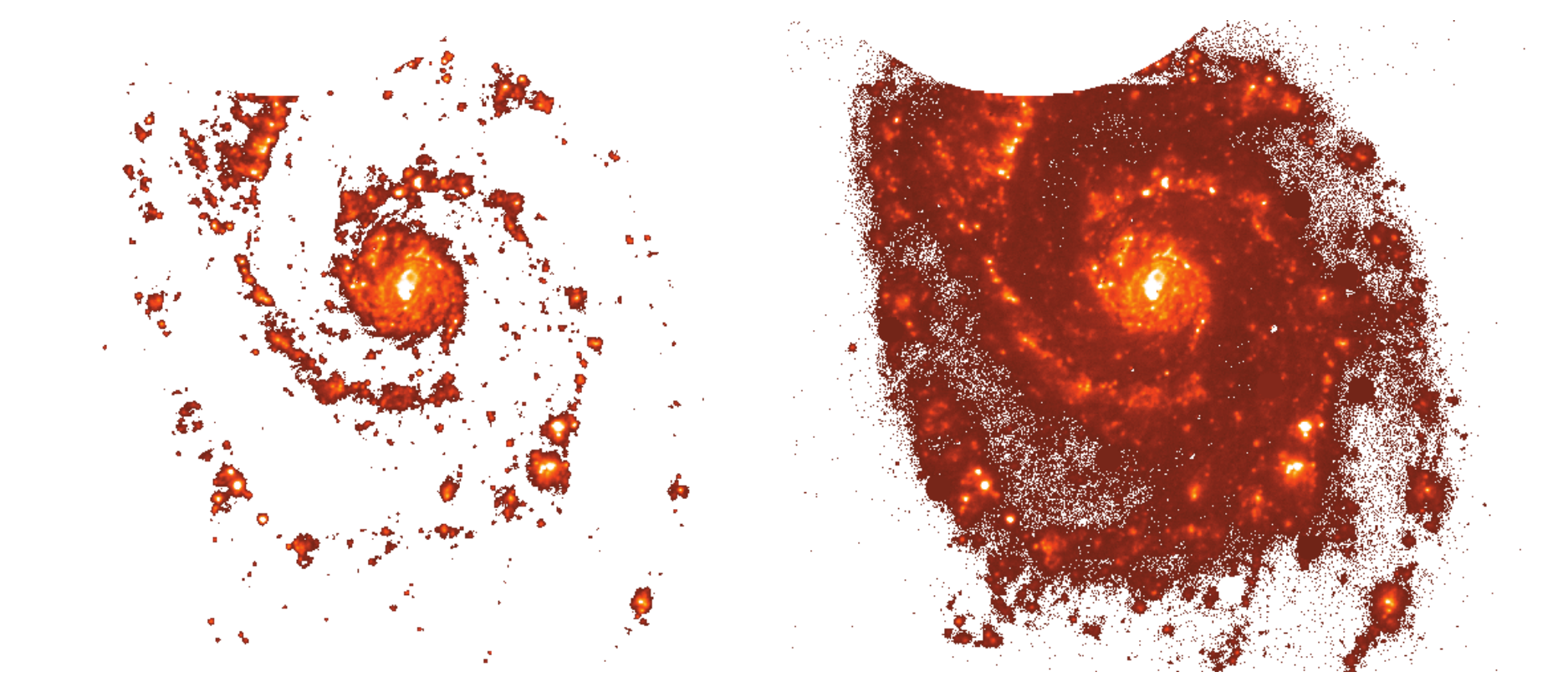}
\end{center}
\caption{\small Both panels show a model 33\,GHz free-free emission image of M\,51 based on an existing H$\alpha$ narrow band image. The left panel shows the emission that would be detected at the 3$\sigma$ level after a $\approx$130\,hr total integration (mosaic with 5\,hr per pointing; $\mathrm{rms} \approx 3\,\mu$Jy\,bm$^{-1}$ with $\theta_{S} = 1\arcsec$, corresponding to a brightness temperature rms of $\approx$4.2\,mK) with the current VLA.  The right panel shows what the ngVLA will deliver after integrating for a total of $\approx$65\,hr (5\,hr per pointing; $\mathrm{rms} \approx 0.76\,\mu$Jy\,bm$^{-1}$ with $\theta_{S} = 1\arcsec$, corresponding to a brightness temperature rms of $\approx$1.0\,mK).  This time estimate additionally takes into account the factor of $\gtrsim$2 larger primary beam for 18\,m antennas. If 12\,m antennas are instead used, the survey speed would increase by another factor of 2.25.  }
\label{fig:m51_model}
\end{figure}

One example of such a free-free emission map is shown in Figure \ref{fig:n5713} for the nearby galaxy NGC\,5713. The left panel illustrates a model 33\,GHz free-free emission based on existing H$\alpha$ narrow band imaging. The H$\alpha$ image was first corrected for contamination from nearby [N{\sc ii}] emission, as well as Galactic and internal dust extinction. The right panel simulates what the ngVLA would observe after 10\,hr using a tapered restoring beam of 1\arcsec, resulting in an rms of $\approx$0.5\,$\mu$Jy\,bm$^{-1}$ ($\approx$0.7\,mK). To reach the same depth with the current VLA would take $\approx$170\,hr.  

As another example, M\,51 (the Whirlpool Galaxy) is one of the most well studied galaxies in the nearby universe, providing a heavily used laboratory for detailed investigations of star formation and the ISM.  With the ngVLA, one could map the entire disk of M\,51 at 30\,GHz down to an rms of $\approx$0.76\,$\mu$Jy\,bm$^{-1}$ with a 1\arcsec~ beam ($\approx$1.0\,mK) in $\approx$65\,hr. A comparison of what can currently be delivered with the VLA for the same integration time per pointing is shown in Figure \ref{fig:m51_model}. To make a map to the same depth using the current VLA would take $\sim$2130\,hr, the same amount of time it to take to roughly survey $\gtrsim$50 galaxies. This is a game-changing step for studies of star formation in the local universe covering a large, heterogeneous set of astrophysical conditions. This statement is independent of the fact that with such observations using the ngVLA, having its wide-bandwidth, a number of RRL's will come for free. The detection of such lines (individually or through stacking), coupled with the observed continuum emission, can be used to quantify physical conditions for the H{\sc ii} regions such as electron temperature. It is without question that the ngVLA will make radio observations a critical component for investigations carried out by the entire astronomical community studying star formation and the ISM of nearby galaxies.  

\begin{figure}[!t]
\begin{center}
\includegraphics[width=\textwidth]{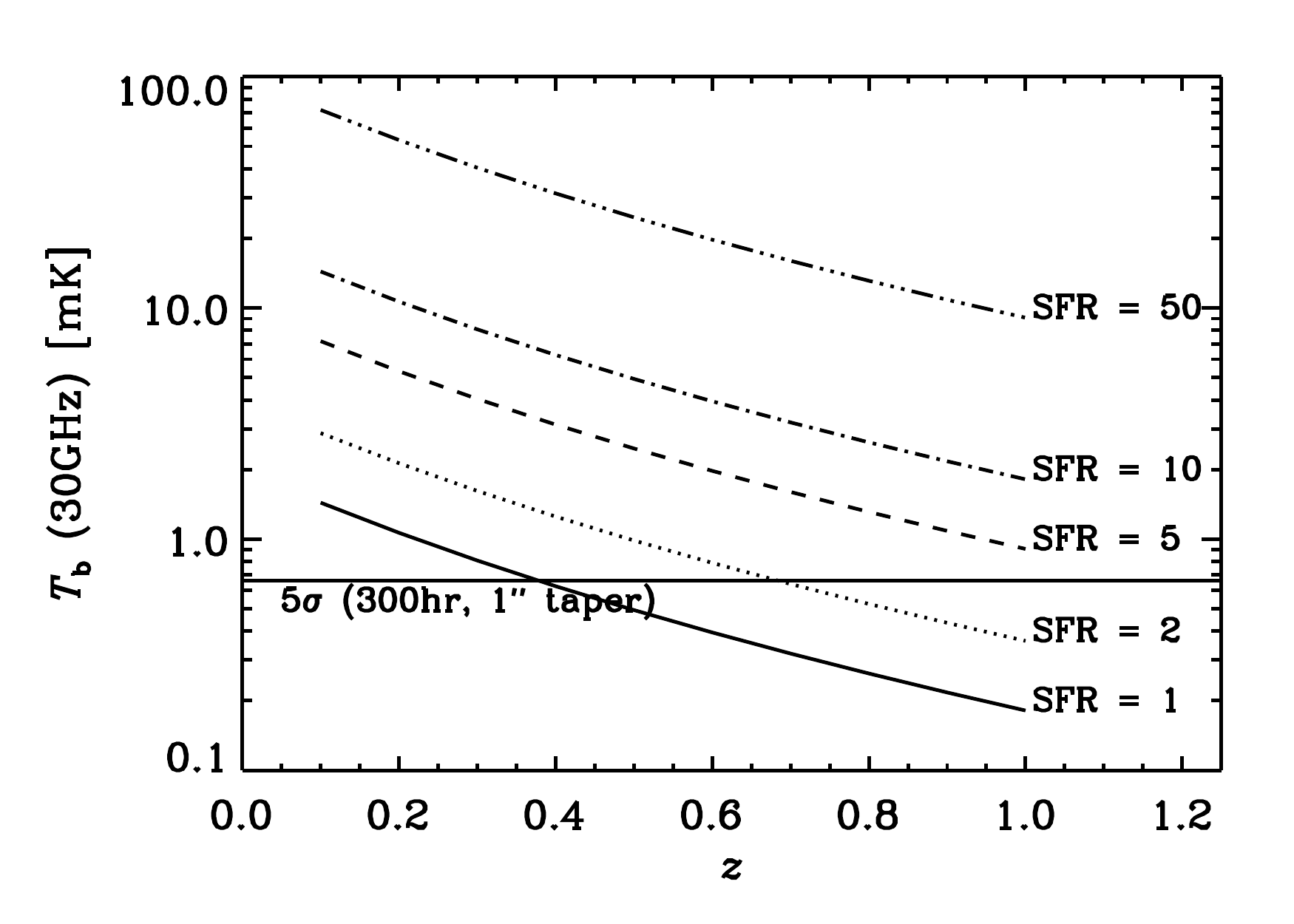}
\end{center}
\caption{\small Observed 30\,GHz brightness temperature vs. redshift indicating the expected brightness temperature of an 8\,kpc disk galaxy forming stars at a rate of 1, 2, 5, 10, and 50\,$M_{\odot}\,\mathrm{yr}^{-1}$.  Given the current sensitivity specifications, by tapering to a 1\arcsec~synthesized beam, the ngVLA will have enough brightness temperature sensitivity to resolve a Milky Way like galaxy forming stars at a rate of a few $M_{\odot}\,\mathrm{yr}^{-1}$ out to $z\sim1$ after a 300\,hr integration. Using the existing VLA, the detection of such a galaxy would take $\sim$1250\,hr.}  
\label{fig:sfr30-vs-z}
\end{figure}

With the increased collecting area and bandwidth of the ngVLA, one would be able to measure and resolve rest-frame $60$\,GHz emission from a Milky Way like galaxy out to $z\sim1$ forming stars at a few $M_{\odot}\,\mathrm{yr}^{-1}$ after a 300\,hr integration. This is illustrated in Figure \ref{fig:sfr30-vs-z} which shows the observed 30\,GHz brightness temperature of 8\,kpc diameter disk galaxies with a range of star formation rates. The 5\,$\sigma$ 30\,GHz brightness temperature rms of the ngVLA tapered to a 1\arcsec~synthesized beam is shown. Accordingly, such observations provide highly robust measurements of star formation rates for comparison with other optical/UV diagnostics to better understand how galaxy extinction evolves with redshift. And, by coupling these higher frequency observations with those at lower radio frequencies one can accurately measure radio spectral indices as a function of redshift to better characterize thermal vs. non-thermal energetics.  

\vspace{6pt}
\noindent
{\em Requirement for a revolution in continuum mapping:}  While the continuum imaging cases described above are a significant improvement over current capabilities, much of this sensitivity gain comes from simply having wider bandwidth receivers that one could conceivably imagine retrofitting on the current VLA. Similar to the technical requirements discussed above for spectroscopic imaging, {\it to make a truly revolutionary step in continuum mapping science of galaxies at these frequencies and at all redshifts will require a substantial amount of collecting area at short (i.e,. $\sim$1\,km) baselines.} This suggestion differs from the notional ngVLA discussed in the cover letter, which only contains $\approx$20\% of the total collecting area in the central 2\,km that is largely comprised of a compact core in the inner 250\,m.  

\vspace{12pt}
\noindent
{\bf The Sunyaev-Zeldovich Effect in Normal Galaxies: The Thermal Pressure Around Galaxies as a New Observable}\\

\vspace{-6pt}
\noindent
The ngVLA offers the opportunity to observe galaxies at wavelengths from $\lambda = 3$~cm to 3~mm with more collecting area and instantaneous bandwidth than has ever been possible. At low redshifts, this frequency range is the sweet spot for investigating the key radio wavelength emission processes: synchrotron, free-free and dust. It also covers the optimum frequency range for observing the Sunyaev-Zeldovich effect, i.e the modification of the CMB by Compton scattering. At ngVLA frequencies, intervening matter produces a reduction in the microwave background which can be used to measure the integrated thermal pressure along the line-of-sight. Most often this process has been observed at low spatial resolution in clusters of galaxies. The ngVLA allows the possibility of making such observations at much higher resolution which would provide more detailed information of the structure of clusters. Furthermore, the potentially large improvement in continuum surface brightness sensitivity means that individual galaxies could also be detected. 

However, such observations will only be possible if a large fraction of the ngVLA (i.e., more than is currently suggested in the cover letter) is placed in a compact configuration optimized for high surface brightness sensitivity. This is because at $\sim$3\,mm, the natural resolution of an array the size of the current VLA D-configuration is $\sim1\arcsec$, a rather high resolution for galaxies. As already discussed in detail above, most spectral line work and even continuum observations also pushes the ngVLA toward such resolution. Also large instantaneous bandwidths are needed for sensitivity, at least $\sim30$--$40$~GHz. The combination of such parameters will allow revolutionary uses of ngVLA for SZ observations. Both the $\nu = 10$--$45$ and $\nu = 70$--$110$~GHz spectral windows would be useful for such observations.

Scaling from the current VLA exposure calculator, if $\sim5\times$ the current VLA collecting area was available on D-configuration baselines with $\Delta \nu = 30$~GHz of bandwidth, then in 10 hours, one would reach rms sensitivities $\sim 6 \times 10^{-3}$~mK. Using the CASA MSMFS imaging algorithm in the 3mm window, one reaches this  sensitivity with $\sim 1\arcsec$ resolution. Assuming a D-configuration-like distribution of baselines, at $\sim 3\arcsec$ resolution, the rms sensitivity for these parameters approaches $\sim 1 \times 10^{-3}$~mK, i.e. a $\mu$K (see EVLA memo 162 for configuration sensitivity details).  

A typical dense cluster observed at arcminute resolution produces $\sim 0.5$~mK signal. The peak surface brightness should increase as the resolution increases, so a quite detailed image of cluster cores should be easily possible. Combined with X-ray imaging, such data would allow the likely complex, 3D distribution of density, temperature and turbulence to be inferred at 1-10 kpc resolution. Smaller objects should also be easily observable such as groups of galaxies and individual D-galaxies. A dense cluster has a pressure, $P / k_B = n T$, $\sim 10^5$~cm$^3$~K. Galaxies might well reach $P / k_B \sim 10^6$ or more in their ISM over perhaps 10~kpc, about $3\arcsec$ at z=0.2. For such pressures, $P / k_B = 10^6$~cm$^{3}$~K, and path lengths of $\sim 10$~kpc the expected signal will be $\sim 2 \times 10^{-2}$~mK.

This integrated thermal pressure at high resolution would be a new observable for galaxies. Combined with other observations, other non-observables could be teased out of these results, such as the pressure from other sources, i.e. magnetic fields and turbulence. The enhancement in pressure due to galaxies being stripped in clusters could be estimated. The pressures is smaller in the disks of nearby star-forming galaxies, which might be measured. Stacking of results for smaller galaxies (or more observing time) might extend the measurements to smaller scales than 10kpc. This discussion just scratches the surface of possible SZ experiments with ngVLA if the array is configured to make such work possible. 

\section{Science Opportunities for the Next Generation VLA with Very Long Baselines}

Very long baseline radio interferometry (VLBI) has proven itself a crucial tool to measure distances, especially to heavily obscured objects, and three dimensional motions. Presently, progress in these areas is limited by a simple lack of sensitivity. As with the other areas described in this white paper, the sensitivity of the proposed ngVLA offers an exciting path forward here. If the ngVLA's collecting area could be harnessed for VLBI observations (see the introduction letter for possible paths) then the ngVLA would be capable of measuring the three dimensional motions of Local Galaxies (and beyond) and the full structure of the Milky Way disk. \\

\noindent {\bf Mapping the Milky Way:} Surprisingly, we know the structure of other galaxies far better than that of the Milky Way. Because we are inside the Milky Way, it has proven very difficult to properly characterize its structure. Beyond just confusing geometry, dust obscures most of the Galaxy at optical and, to some extent, infrared wavelengths. As a result, distances to objects beyond the extended Solar Neighborhood are often quite uncertain. Thus, we only have an ``educated guess'' that the Milky Way is a barred Sb or Sc galaxy, and even the number of spiral arms (2 or 4) is actively debated.

\begin{figure}[!t]
\centering
\begin{minipage}{0.4\textwidth}
\centering
\includegraphics[height=7cm]{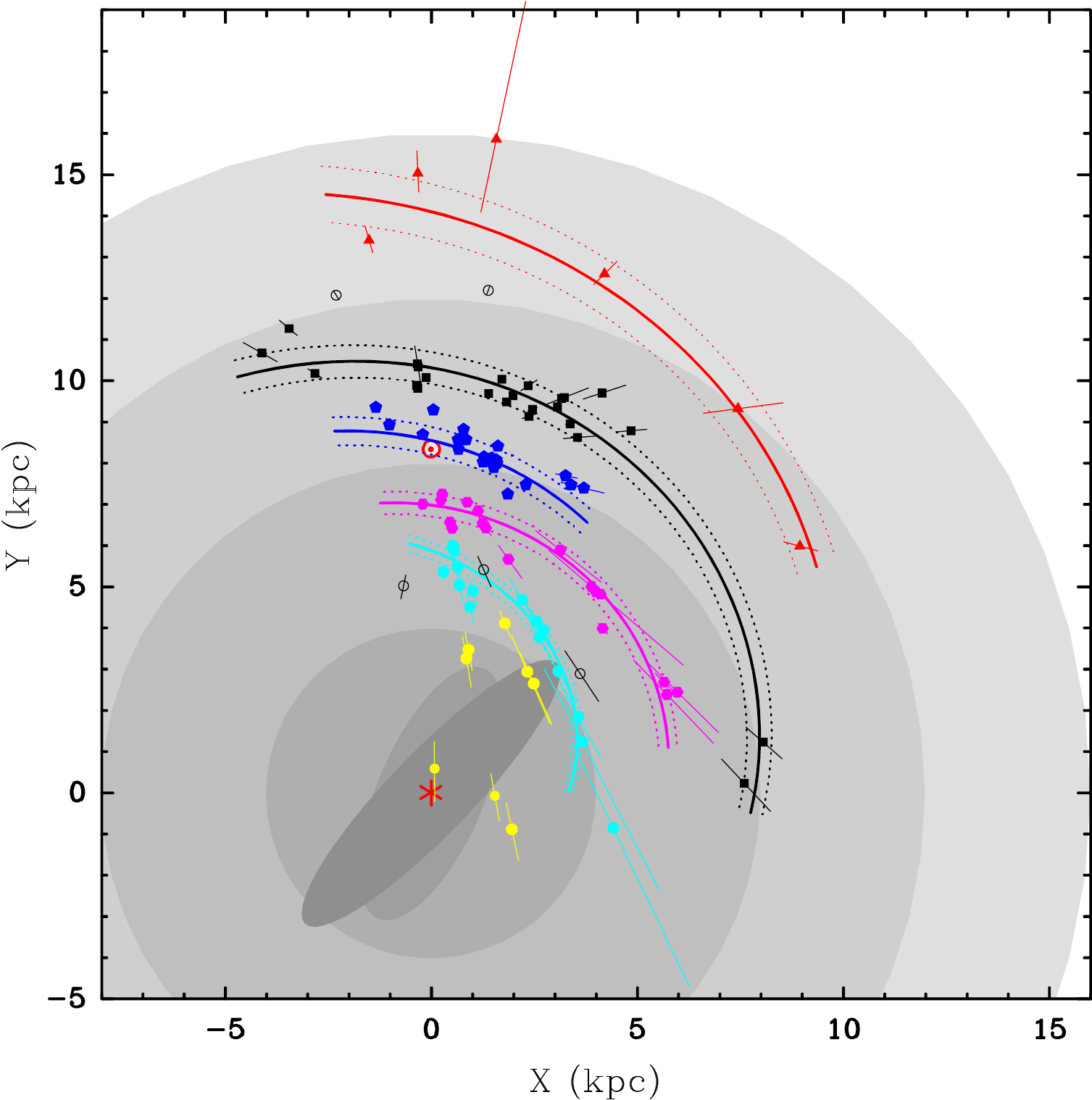}
\end{minipage}\hfill
\begin{minipage}{0.425\textwidth}
\centering
\caption{\small Schematic view of the Milky Way showing the locations of high-mass star forming regions with measured trigonometric parallaxes from VLBA maser observations. Most of the measured distances are still near the Sun (red Sun symbol at 0,~8.34) and the region beyond the Galactic center (red asterisk at 0,0) is largely unknown because of the limited sensitivity of the current VLBA. Harnessing the very large proposed collecting area of the ngVLA for very long baseline applications could improve this situation, allowing full mapping of the structure of the plane of the Milky Way. Such mapping would be a powerful complement to GAIA, which will measure exquisite parallaxes for a billion stars but will not be able to penetrate the heavy extinction obscuring the far side of the Milky Way disk.
\label{fig:parallaxes}}
\end{minipage}
\end{figure}

Using the current VLBA, the BeSSeL Survey is mapping the spiral structure of the near side of the Milky Way. However, this effort is limited by the sensitivity of the VLBA to distances of only about 10 kpc (see Fig.~\ref{fig:parallaxes}). Thus this effort can only reach very slightly beyond the Milky Way center, leaving the structure of significant portion of the far side of our Galaxy essentially unknown.

Beyond the intrinsic interest in understanding our home, distances to star forming regions across our Galaxy are essential to understand its full spiral structure, e.g., to connect nearby spiral arm segments in the fourth Galactic quadrant with their continuation in the first quadrant, and so to place the Milky Way within the broader population of galaxies.  

The large proposed collecting area of the ngVLA offers the prospect to extend parallax measurements of Milky Way masers to the far side of the galaxy. If a large part of this collecting area could be harnessed for long baseline measurements, one could reasonably expect to weak masers to measure distances to better than 10\% accuracy at distances of 20 kpc. Note that this goal is extremely complementary to the GAIA mission. GAIA will revolutionize our knowledge of the Milky Way along relatively low extinction sight lines but cannot penetrate the enormous extinctions associated with the plane of the Milky Way. High precision radio parallaxes will remain the best way to map out the structure of the plane of the Milky Way itself. By combining these two approaches, the next decade will see our knowledge of the three dimensional structure of the Milky Way come into its own. The ngVLA can play a key role in this that is unlikely to be accomplished using any other technique in the meantime.\\

\noindent {\bf Local Group Cosmology}: The distribution of dark matter in galaxies and groups of galaxies is one of the major problems in observational cosmology. The Local Group provides the nearest and best system for detailed study.  So far, attempts to ``weigh'' the Milky Way and the Andromeda galaxy have resulted in inconclusive values. The major problem is that most studies work with only one-dimensional (radial) velocity components, introducing significant ambiguities and requiring inference based on small sample sizes and/or unknown biases from  non-isotropic velocity distributions. The path to reliable mass distributions is three dimensional velocity measurements that combine radial velocities with proper motions.

Understanding both the history and fate of the Local Group also requires detailed knowledge of the three dimensional space motion of the Local Group, specifically the motion of Andromeda with respect to the Milky Way. If Andromeda has a negligible proper motion (with respect to the Milky Way), then these two dominant Local Group members will collide and merge over the next few Gyr.   However, if Andromeda has a sizable proper motion, of order 100 \kms, as suggested in the literature, the stellar disks would not collide and the two galaxies would pass by each other and not merge on this time scale.

We are currently poised to make dramatic progress in understanding the dynamics, and hence dark matter mass distribution, of the Local Group. The proper motion of the Andromeda galaxy is the key measurement for this.  There are two ways to measure Andromeda's proper motion: 1) directly, via its AGN (M31*) or 2) indirectly, by measuring the motions of newly discovered water masers in Andomeda's disk. M31* is weak ($\sim30$ \uJy) and variable, making it extremely difficult to detect with high signal-to-noise with current VLBI arrays.   However, incorporating the ngVLA's large collecting area into very long baselines would make the required measurements straightforward. Over $\approx10$ years, we could measure the motion of M31* with accuracies of $\sim0.1$~\uasy\ and thus measure the three-dimensional velocities of Andromeda and other galaxies in the Local Group with uncertainties less than 1 \kms .  For comparison, current VLBI measurements of water masers in M~33 and IC10 have achieved $\approx5$ \uasy\ accuracy over time baselines of less than 5 years (and optical proper motion measurements, e.g., of the LMC have comparable accuracy), so this would represent more than an order of magnitude increase in accuracy over present day capabilities. Indeed, with this accuracy, significant measurements of galaxy motions would be possible in other groups of galaxy out to the Virgo Cluster.\\

\noindent {\bf Technical Requirements:} The key technical requirement to enable these key science goals is that the ngVLA include VLBI capabilities as part of its baseline design. As discussed in the introduction letter, several schemes are plausible given that these applications focus less on high fidelity imaging than precise astrometry and so have less stringent $u-v$ coverage requirements than some of the imaging applications discussed above. In any case, the high profile, important nature of this science strongly argues that this capability not be left until after the array construction but be considered as a core part of ngVLA design.

\end{document}